\documentclass[aps,twocolumn,showpacs,reprint,amsmath,amssymb,pra,superscriptaddress,floatfix,longbibliography]{revtex4-1}
\usepackage{braket}
\usepackage{amsmath}
\usepackage{amssymb}
\usepackage{mathtools}
\usepackage{graphicx}% Include figure files
\usepackage{dcolumn}% Align table columns on decimal point
\usepackage{bm}% bold math
\usepackage{multirow}
\usepackage{leftidx}
\usepackage{color}
\usepackage{float}
\usepackage{hyperref}

\usepackage{amsfonts}
\usepackage[german,english]{babel}

\begin{document}

\preprint{APS/123-QED}

\title{Test of Nuclear Decay Rate Variation due to an Antineutrino Flux}% Force line breaks with \\
%\thanks{A footnote to the article title}%

\author{Shih-Chieh Liu}
\author{David Koltick}
\email[]{Corresponding Author: koltick@purdue.edu}
\author{Haoyu Wang}
\affiliation{Department of Physics and Astronomy, Purdue University, West Lafayette, IN 47906}

\author{Jonathan Nistor}
\author{Jordan Heim}
\author{Thomas Ward}
\affiliation{TechSource Inc, Los Alamos, NM 87544}

\date{\today}% It is always \today, today,
             %  but any date may be explicitly specified

\begin{abstract}
\noindent
\textbf{Background:} Unexplained variations of the decay rate parameter for weak interaction decays such as $\beta^{\pm}$-decay, electron capture, as well as strong interaction $\alpha$-decay have been reported. Because these variations have been presented by a number of groups, at various locations, using various types of detectors, different isotopes, and over extended periods of time, some researchers have interpreted the source of these variations as not from ambient environmental factors such as temperature, pressure, and humidity but via an unexplained fundamental interaction.\\ 
\textbf{Purpose:}  To review the state of decay rate parameter variations experiments and place the reported results into a common comparable context using a temporal cross section. Then to make decay parameter measurements as a function of time at the level of 10$^{-5}$ in the presences of an antineutrino flux with an on-off cycle time of $\sim$30 days. This level of precision requires a detailed understanding of both systematic and statistical errors, otherwise, systematic errors in the measurement may mimic the decay events of fundamental interactions. Two weak interaction decays, one via electron capture and the other via $\beta^-$ decay were selected because the final state and the time reverse state each contain a neutrino and anti-neutrino, covering arguments that the anti-neutrino flux may interact differently or not at all in one of the cases. \\ 
\textbf{Methods:} The experiment searched for variation of the $^{54}_{25}$Mn, e$^-$ capture decay rate parameter and $^{137}_{~55}$Cs, $\beta^-$ decay rate parameter both to a level of precision of 1 part in $\sim 10^{5}$ by comparing the difference between the decay rate,  in the presence of an antineutrino flux $\sim 3 \times10^{12}$ $\overline{\nu} \, cm^{-2} \, sec^{-1}$ and no flux measurements. The experiment was located 6.5 meters from the reactor core of the High Flux Isotope Reactor(HFIR) in Oak Ridge National Laboratory.\\
\textbf{Results:} The $\gamma$ spectra from both decays were collected and analyzed independently. The measured variation in the decay rate parameters are found to be $\delta \lambda/ \lambda = (0.034\pm 1.38)\times 10^{-5}$ for $^{54}_{25}$Mn and $\delta \lambda/ \lambda = (0.67\pm 1.56)\times 10^{-5}$ for $^{137}_{~55}$Cs.  These results are consistent with no measurable decay rate parameter variation due to an antineutrino flux, yielding a $68\%$ confidence level upper limit sensitivity for $^{54}_{25}$Mn  $\delta \lambda/ \lambda \leq 1.31\times10^{-5}$ or $\sigma \leq 1.29\times10^{-25}\,cm^{2}$ in cross section and  for $^{137}_{~55}$Cs $\delta \lambda/ \lambda \leq 2.23\times10^{-5}$ or $\sigma \leq 5.69\times10^{-27}\,cm^{2}$. \\
\textbf{Conclusions:} The cross-section upper limit obtained in these null or no observable effect measurements are $\sim 10^{4}$ times more sensitive than past experiments reporting positive results in $^{54}$Mn and $^{137}$Cs.

%\begin{description}
%\item[Usage]
%Secondary publications and information retrieval purposes.
%\item[PACS numbers]
%May be entered using the \verb+\pacs{#1}+ command.
%\item[Structure]
%You may use the \texttt{description} environment to structure your abstract;
%use the optional argument of the \verb+\item+ command to give the category of each item. 
%\end{description}
\end{abstract}

%\pacs{Valid PACS appear here}% PACS, the Physics and Astronomy
                             % Classification Scheme.
%\keywords{Suggested keywords}%Use showkeys class option if keyword
                              %display desired
\maketitle

%\tableofcontents

%\section{\label{sec:level1}First-level heading}
\section{Introduction}
Unexplained variations of the decay rate parameter for weak interaction decays such as $\beta^{\pm}$-decay, and electron capture \cite{ALBURGER1986168,Falkenberg200132,Falkenberg200241,Jenkins201350,Jenkins200942,Jenkins201281,OKeefe2013,Schrader20101583,Schrader2016202,Shnoll199810,Siegert19981397,Shnoll200043,Sturrock2011,Sturrock201362,Sturrock201447,Sturrock2012755,Sturrock201218,Sturrock20168} as well as strong interaction $\alpha$-decay\cite{Jenkins200942,Siegert19981397} have been reported. Because the variation of the decay rate parameter has been presented by a number of groups, located at various locations, using various types of detectors, different isotopes, and over extended periods of time, some researchers have interpreted the source of these variations as not from ambient environmental factors such as temperature, pressure, and humidity but via an unexplained fundamental interaction. Some results show a correlation between an annual periodicity of the decay rate parameter variation, and the variable distance of the Earth from the Sun. The annual variation of the Earth-Sun distance causes a $\sim7\%$ variation of the total neutrino flux on the Earth. This flux variation as the source of the decay rate parameter variation is motivated by the large neutrino flux, $6.5\times10^{10}\,\nu\, cm^{-2}sec^{-1}$, on the Earth dominated by solar fusion. Also, some researchers suggest that decay rate parameters are affected by solar activity such as solar flares\cite{Bellotti2013116,Fischbach2009,Jenkins2009407,Krause201251}.\\
\indent
However, these conclusions are controversial. This research is focused on the possibilities that the reported variations are an extension of weak interactions. Conventional weak interaction neutrino-nucleon cross sections are 20 orders of magnitude smaller than the reported strong interaction level cross-sections being observed in these experiments, if caused by neutrinos. The conventional neutrino interaction cross-section per nucleon ($\nu + n \rightarrow p^{+} + e^{-}$) is \cite{RevModPhys.84.1307}
\begin{equation}
\begin{split}
\sigma_{weak}&\sim \dfrac{4G_{F}^{2}E_{\nu}^{2}(\hbar c)^{2}}{\pi} \\
&\sim 9 \times 10^{-44}\,cm^{2} \left( \dfrac{E_{\nu}}{1\,MeV} \right)^{2} \\
\end{split}
\label{weak_cross}
\end{equation}
where $G_{F}$ is the Fermi constant, and $E_{\nu}$ is the neutrino energy. Because antineutrinos interact with protons ($\overline{\nu}  + p^{+} \rightarrow n + e^{+}$), there is no significant difference between neutrino, and antineutrino interaction cross-section on target nuclei. That is neutron, and proton number are similar $\sim A/2$. If a typical neutrino or antineutrino energy is considered to be $\sim\,1$ MeV, Eq.\ref{weak_cross} leads to a nucleon cross section of
\begin{equation}
\sigma_{weak} \sim N^{\beta} \times 9 \times 10^{-44} cm^{2}
\end{equation}
where $N \sim A/2$ is the number of neutrons or protons in the nucleus. The antineutrino cross section is proportional to $N^{\beta}$. $\beta = 2$ if the antineutrino scattering is coherent from the nucleus, and $\beta = 1$ if the scattering is incoherent. The largest possible conventional cross section occurs if the scattering is coherent.\\
\indent
The goal of this experiment is to maximize the sensitivity to the decay parameter variation caused by a possible extension to the weak interactions. The disadvantage of solar neutrinos as a test source is the low flux variation $\sim 10^{9}\, \nu \, cm^{-2} \, sec^{-1}$, which demands long measurement time. In addition, environmental influences, a source of systematic error, follow this same time cycle.  Antineutrinos, on the other hand, can be generated from a nuclear reactor having a stable antineutrino flux ($\sim 3\times10^{12}$ $\overline{\nu} \, cm^{-2} \, sec^{-1}$), and much larger flux variations due to the reactor off cycles ($\sim 0 \, \nu \, cm^{-2} \, sec^{-1}$)\cite{Nuclear}. For this reason, the experiment was performed at the High Flux Isotope Reactor (HFIR) at Oak Ridge National Laboratory (ORNL). The HFIR provides reactor-on, and reactor-off cycles of similar duration. \\
\indent
Specifically, the first set of experiments recorded the $\gamma$ spectra from $^{54}_{25}$Mn electron capture decay
\begin{equation}
\begin{split}
_{25}^{54}Mn (3^{+}) + e^{-}\to\,&{}_{24}^{54} Cr^{*}(2^{+})+{\nu _e}\\
&\rightarrow \,_{24}^{54}Cr(0^{+})  + \gamma\,(834.8\,keV).\\
\end{split}
\end{equation}
Even though the maximum possible weak cross section, coherent, for $^{54}$Mn($Z=25$ with $\beta=2$) is expected to be at the level of
\begin{equation}
\sigma_{weak} \sim 6 \times10^{-41}\,cm^{2},
\end{equation}
the cross-section sensitivity to decay parameter variation obtainable at HFIR is $\sim$10$^{4}$ more sensitive than the reported positive results yielding strong interaction like cross-sections. The use of antineutrinos in no way invalids comparison between this, and previous experimental effects based on neutrinos. $^{54}$Mn was specially selected because the basic interaction involves a proton ($p^{+}+e^{-}\rightarrow n +\nu$) matching the inverse $\beta^{-}$-decay reaction on neutrons caused by neutrinos.\\
\indent
Similarly, for completeness with a proton in the final state, a second set of experiments recorded the $\gamma$ spectra from $^{137}_{~55}$Cs $\beta^-$ decay
\begin{equation}
\begin{split}
_{~55}^{137}Cs (7/2^{+}) \to\,  &{}_{~56}^{137} Ba^{*}(11/2^{-}) +\beta^-+ {\bar \nu _e}\\
&\rightarrow \, {}_{~56}^{137}Ba(3/2^{+})  + \gamma\,(661.660\,keV).\\
\end{split}
\end{equation}
In this case if weak coherent interactions were to occur involving the neutrons the expected cross section level would be, 
\begin{equation}
\sigma_{weak} \sim 6 \times10^{-40}\,cm^{2}.
\end{equation}
\section{Historical: Variation of Radioactive Decay Rate Parameters}
Oscillation variations of the radioactive decay rate parameters have been investigated for several decades. A number of groups such as Alburger, Falkenberg, Veprev, and Jenkins $et.\,al.$ observed time-dependent decay rate parameter variations\cite{ALBURGER1986168,Falkenberg200132,Falkenberg200241,Jenkins201350,Jenkins2009407,Jenkins200942,Jenkins201281,Veprev201226}, believed not due to variations in ambient environmental factors such as temperature, pressure, and humidity. Some results show a correlation between an annual periodicity of the decay rate parameter, and the variable distance of the Earth from the Sun. These results motivate some researchers to conclude the solar neutrino flux variations cause the decay rate parameter variations\cite{Jenkins201350,Jenkins2009407,Jenkins200942,Jenkins201281}.\\
\indent
However, the correlation of the radioactive decay rate parameter variation, and solar neutrino flux has been challenged by Kossert, Semkow, Meijer, Bruhn , Schrader, and Bellotti $et.\,al.$\cite{Bruhn200228,Kossert201433,Kossert201518,deMeijer2011320,Schrader20101583,Schrader2016202,Semkow2009415}. These "null evidence" experimental references attribute the decay rate parameter variation to ambient environmental factors or instrumental error, instead of the solar neutrino flux variation. \\
\indent
In this section, a summary of some key results will be discussed. The results are sorted by isotopes, and published date. A key comment, and a short summary table are presented at the beginning of each discussion. "Positive" represents those observations reporting time-dependent decay rate parameters, whereas "Negative" are those showing no variation results. The size of the observed effect or the experimental sensitivity($\delta \lambda/\lambda$) in the case of null results are also displayed in the short summary table in each discussion. Where as Table~\ref{limit_yes}, and Table~\ref{limit_no} display the results of the full literature review, and cross-section sensitivities, Eq.~\ref{crosssection1}, for each reported isotope.
\subsection*{$^{32}$Si and $^{36}$Cl} 
\begin{table}[ht!]
\begin{center}
\caption{Summary of the $^{32}$Si, and $^{36}$Cl results.}
\begin{tabular}{cccc}
\hline \hline
\multirow{2}{*}{Source}& \multirow{2}{*}{Reference} & Sensitivity & \multirow{2}{*}{Variation} \\ 
 &  & (Size of Effect) &  \\ 
\hline 
$^{32}$Si/$^{36}$Cl &Alburger $et.\,al.$  & $5\times10^{-3}$&Positive\\
$^{36}$Cl &Kossert $et.\,al.$  & $4\times10^{-4}$&Negative \\ 
$^{32}$Si &Semkow $et.\,al.$  & $1\times10^{-3}$&Negative \\
\hline \hline
\end{tabular}
\label{T_32Si}
\end{center}
\end{table}
\textbf{Alburger} $et.\,al.$\cite{ALBURGER1986168} reported the first observation of decay rate parameter variations in 1986. Alburger worked on the half life measurement of $^{32}$Si with a gas proportional detector over the period 1982 through 1986 at Brookhaven National Laboratory(BNL). They unexpectedly observed small periodic annual deviations of the data points from an exponential decay. The authors reported that temperature, and humidity \textit{"can not fully account"} for observed ratio $^{32}$Si$/^{36}$Cl decay rate variation during their measurement, as shown in Table~\ref{T_32Si}. However, they do not have complete environment records, to backup these claims.\\
\indent
\textbf{Kossert} $et.\,al.$\cite{Kossert201433} reported no decay rate parameter variations of $^{36}$Cl measured using a custom-built triple-to-double coincident ratio detector (TDCR) at Physikalisch-Technische Bundesanstalt (PTB). TDCR is an optical chamber with three photomultiplier tubes surrounding a liquid scintillation detector. The sample is placed in the center. A triple coincidence detector is much less sensitive to ambient environmental factors. Kossert measured much smaller variation of $^{36}$Cl as shown in Table \ref{T_32Si} than observed by Alburger, but Kossert attributes the small variation to instrumental effects, instead of variations of the decay rate parameter. This conclusion is much more reliable than Alburger's because Alburger used a single gas proportional detector which is sensitive to environmental variations.\\
\indent
\textbf{Semkow}\cite{Semkow2009415} has written a review concerning decay rate parameter variations of $^{32}$Si from Alburger's results. Semkow explained the variations of $^{32}$Si by the change of temperature, causing the air density in the space between the source and the gas proportional detector to change, reducing and increasing the count rate. The higher temperature in the summer causes lower air density in the space between the source and the detector, and resulting in less absorption of the lower energy $\beta$ particles in the air. Thus, the gas proportional detector collects more $\beta$ particles at a higher temperature, generating a higher counting rate. The resulting limit is given in Table~\ref{T_32Si}.
\subsection*{$^{152}$Eu, $^{154}$Eu, and $^{155}$Eu } 
\begin{table}[ht!]
\begin{center}
%\setstretch{1.25}
\caption{Summary of the $^{152}$Eu, $^{154}$Eu, and $^{155}$Eu results.}
\begin{tabular}{cccc}
\hline \hline
Source & Reference & Sensitivity & Variation \\ 
\hline 
 $^{152}$Eu, $^{154}$Eu,$^{155}$Eu  &Siegert $et.\,al.$  & $5\times10^{-4}$&Negative\\
 $^{152}$Eu &Meijer $et.\,al.$  & $1.4\times10^{-4}$&Negative \\
\hline \hline
\end{tabular}
\label{Eu}
\end{center}
\end{table}
\indent
\textbf{Siegert} $et.\,al.$\cite{Siegert19981397} studied the multi $\beta$-decay modes of Eu isotopes. Siegert used the strong interaction $\alpha$-particle $^{226}$Ra decay as a reference to determine the half life of the weak interaction $\beta$ decays of $^{152}$Eu, $^{154}$Eu, and $^{155}$Eu, measured using two different kinds of detector systems; an ion chamber, and a solid state detector(Ge, Li) at Physikalisch-Technische Bundesanstalt  (PTB). $^{152}$Eu decays to $^{152}$Gd  by electron capture with 72.1$\%$ branching ratio, and to $^{152}$Sm by $\beta^{-}$-decay with branching ratio 27.9$\%$. Siegert reported that the oscillations of $^{226}$Ra have a maximum positive deviation in February, and minimum deviation in August. Siegert observed oscillations in $^{226}$Ra, as well as in the other isotopes, but explains the effect as follows:
\begin{quotation}
\textit{"A discharge effect on the charge collecting capacitor, the cables, and the insulator to the ionization chamber electrode caused by background radioactivity such as radon, and daughter products which are known to show seasonal concentration changes."}
\end{quotation}
Siegert concludes the oscillations are proportional to the ionization current. If the oscillations were due to solar neutrinos interacting with isotopes via the weak interactions, the $^{226}$Ra strong interaction decay oscillations should not depend on the ionization current. With these considerations, Siegert's observation are considered to be upper limits, as shown in Table~\ref{Eu}.\\
\indent
\textbf{Meijer} $et.\,al.$\cite{deMeijer2011320} used reactor antineutrinos as a source. Meijer reported null evidence for the decay rate variation of $^{152}$Eu using reactor antineutrinos, as shown in Table~\ref{Eu}. If the solar neutrino variations cause the decay rate parameter variation, Meijer should have observed a stronger effect compared to Siegert due to the factor of 10 higher antineutrino flux variation from the reactor cycling. No effect was observed.
\subsection*{$^{3}$H}
\begin{table}[ht!]
\begin{center}
%\setstretch{1.25}
\caption{Summary of the $^{3}$H results.}
\begin{tabular}{cccc}
\hline \hline
\multirow{2}{*}{Source}& \multirow{2}{*}{Reference} & Sensitivity & \multirow{2}{*}{Variation} \\ 
 &  & (Size of Effect) &  \\ 
\hline 
$^{3}$H &Falkenberg   & $3.7\times 10^{-3}$ &Positive\\
$^{3}$H &Bruhn   & $2\times10^{-3}$&Negative \\ 
$^{3}$H &Veprev $et.\,al.$ & $2\times10^{-1}$&Positive\\
\hline \hline
\end{tabular}
\label{H3}
\end{center}
\end{table}
\textbf{Falkenberg}\cite{Falkenberg200132} is the first one to put forward the hypothesis that the variations of the $\beta$-decay rate parameters are due to the solar neutrino flux variations. Falkenberg measured the radioactive $\beta$-decay rate parameter of tritium by a photodiode detector from 1980 to 1982, as shown in Table~\ref{H3}. To determine the significance of the data's periodic deviation, the residuals were first fit to a single periodic function. Falkenberg calculated the residuals as the differences between an aperiodic exponential form, and the data. Then, he included a cosine function in the fit with a period of 365 days in order to account for the variation away from the aperiodic function. Falkenberg concluded:
\begin{quotation}
\textit{"There is a positive correlation between the periodically changing solar neutrino flux, and the $\beta$-decay of 
tritium."}
\end{quotation}

\textbf{Bruhn}\cite{Bruhn200228} re-analyzed Falkenberg's data, and criticized Falkenberg for not making corrections for any background effects in his tritium decay rate measurements. In addition, Bruhn concludes Falkenberg's results are not sufficient for deducing a correlation between the tritium decay rate, and the orbital motion of the Earth because neither the period nor amplitude of the deviation coincides with the orbital motion of the Earth. Bruhn concludes:
\begin{quotation}
\textit{"By taking the deviation of the measurement data with respect to the optimal solution instead of the true solution} (in the fits) \textit{E.D. Falkenberg cannot separate any (hypothetical) additional background effects in his data from the true solution."}
\end{quotation}
Bruhn's analysis of Falkenberg's data can be used to estimate a limit on the decay parameter variations in $^{3}$H. Both results are included in this review, as shown in Table~\ref{H3}.\\
\indent
\textbf{Veprev} $et.\,al.$\cite{Veprev201226} measured the high-energy region of the tritium beta decay spectrum using a liquid scintillation detector system viewed by three photomultipliers. Veprev reported decay rate parameter oscillations which coincide with the solar neutrino flux variation distance from the Earth to the Sun, as shown in Table~\ref{H3}. Veprev concludes that the periodicity of the tritium decay rate parameter variations is due to the interactions of the tritium nuclei with solar neutrinos.
\subsection*{$^{54}$Mn}
\begin{table}[ht!]
\begin{center}
%\setstretch{1.25}
\caption{Summary of the $^{54}$Mn results.}
\begin{tabular}{cccc}
\hline \hline
\multirow{2}{*}{Source}& \multirow{2}{*}{Reference} & Sensitivity & \multirow{2}{*}{Variation} \\ 
 &  & (Size of Effect) &  \\ 
\hline 
$^{54}$Mn &Jenkins $et.\,al.$  & $1\times 10^{-3}$ &Positive\\
$^{54}$Mn &Meijer $et.\,al.$ & $4\times10^{-4}$&Negative\\
\hline \hline
\end{tabular}
\label{54Mn}
\end{center}
\end{table}
\indent
\textbf{Jenkins} $et.\,al.$\cite{Jenkins201281,Jenkins2009407} was the first to conclude there is a decay rate parameter variation due to the variable distance of the Earth from the Sun. The results are shown in Table~\ref{54Mn}. In addition, Jenkins $et.\,al.$ reported the detection of a significant decrease in the decay rate parameter of $^{54}$Mn during a strong solar flare at the end of 2006. Jenkins measured the count rate of $^{54}$Mn, and compared it with the Solar X-ray data. The deviation is clearly visible on 12/12/06, through 12/17/06, which was coincident with a severe solar storm. Jenkins attributed the annual oscillations observed in the data to the variations in solar neutrino flux due to the annual variation in the distance between the Sun, and the Earth. \\
\indent
\textbf{Meijer} $et.\,al.$\cite{deMeijer2011320} used reactor antineutrinos as a source. Meijer reported null evidence of the $^{54}$Mn electron capture decay rate parameter to vary due to an antineutrino flux with improved sensitivity relative to Jenkins'\cite{Jenkins201281,Jenkins2009407}, as shown in Table~\ref{54Mn}. The experiments were conducted comparing the $\gamma$-ray count rate during reactor on, and off periods at an antineutrino flux of $\sim 5 \times10^{10}\,\overline{\nu}\,cm^{-2}sec^{-1}$. The results showed no variations of the $^{54}$Mn decay rate parameter. This challenges Jenkin's conclusions because the solar neutrino flux on the Earth varies only $\sim 7 \%$ ($4.6\times10^{9}\,\nu \,cm^{-2}sec^{-1}$). Hence, Meijer should have observed an effect more than 10 times larger than Jenkin's, if Jenkin's hypothesis, were correct.\\

\subsection*{$^{137}$Cs}
\begin{table}[ht!]
\begin{center}
%\setstretch{1.25}
\caption{Summary of the $^{137}$Cs results.}
\begin{tabular}{cccc}
\hline \hline
Source & Reference & Sensitivity & Variation \\ 
\hline 
$^{137}$Cs &Schrader & $4.6\times 10^{-4}$ &Negative\\
$^{137}$Cs &Bellotti $et.\,al.$ & $8.5\times10^{-5}$&Negative\\
\hline \hline
\end{tabular}
\label{137Cs}
\end{center}
\end{table}
\indent
\textbf{Schrader} $et.\,al.$\cite{Schrader20101583} observed variations in the decay rate measurements of $^{137}$Cs using an ionization chamber at Physikalisch-Technische Bundesanstalt (PTB). Schrader concludes the small yearly variations are from the measuring electronics, instead of decay rate parameter variations. The results are shown in Table~\ref{137Cs}.\\
\indent
\textbf{Bellotti} $et.\,al.$\cite{Bellotti2013116} measured the decay rate of $^{137}$Cs radioactive source  using a NaI scintillation detector, and a Ge semiconductor detector. The results are shown in Table~\ref{137Cs}. No significant yearly deviation from the expectations was measured. In addition, the data exhibited no decay rate parameter variations in the presence of the two solar flares of the year 2011, and 2012.
\subsection*{$^{226}$Ra} 
\begin{table}[ht!]
\begin{center}
\caption{Summary of the $^{226}$Ra results.}
\begin{tabular}{cccc}
\hline \hline
\multirow{2}{*}{Source}& \multirow{2}{*}{Reference} & Sensitivity & \multirow{2}{*}{Variation} \\ 
 &  & (Size of Effect) &  \\ 
\hline 
 $^{226}$Ra &Siegert $et.\,al.$  & $1\times10^{-3}$&Negative\\
 $^{226}$Ra &Jenkins $et.\,al.$  & $2\times10^{-3}$&Positive\\
 %$^{226}$Ra &Semkow $et.\,al.$  & $5\times10^{-3}$&Negative \\
\hline \hline
\end{tabular}
\label{Ra}
\end{center}
\end{table}
\indent
\textbf{Siegert} $et.\,al.$\cite{Siegert19981397} used $^{226}$Ra, having a strong interaction $\alpha$-particle decay, to study decay parameter variations. Siegert measured $^{226}$Ra decay rate with an ionization chamber at the PTB. Siegert reported that the oscillations of $^{226}$Ra have a maximum positive deviation in February, and minimum deviation in August. He accounted for these oscillations as due to seasonal environmental variations. This conclusion meets the expectation of a null result if extensions to weak interactions were the source of parameter variations. The results are shown in Table~\ref{Ra}.\\
\indent
\textbf{Jenkins}\cite{Jenkins200942} re-analyzed Siegert's $^{226}$Ra decay rate data and showed the observed variations have a correlation to the inverse squared distance between the Earth, and the Sun. The results are shown in Table~\ref{Ra}.

\subsection{Summary of Literature}
The literature does not report a consistent picture of decay rate parameter variations. Most of the references have used the Sun as a source of the rate variation through neutrino interactions.   However, there is no discussion relating the reported effect or null effect in terms of a cross section sensitivity. For this reason, the results are not directly comparable, one to another, without such a framework. All authors except one report only the size of the effect relative to the decay rate. This work attempts to place previous results into a unified framework expressed as an interaction cross section due to the neutrino or anti-neutrino flux observed in the experiment. Furthermore, cross-section sensitivity provides an opportunity to examine further how to proceed in studies of possible decay rate parameter variations of radioactive isotopes.
\begin{table*}
\caption[Summary of experiments that conclude time-dependence of radioactive decay rate parameters.]{Summary of experiments that conclude \textbf{Time-dependence} of radioactive decay rate parameters. Estimation of the interaction cross section is based on Eq.~\ref{crosssection1} by inputting the associated size of effect, mean lifetime as well as neutrino or antineutrino flux from each reference. The variation of the solar neutrino flux is taken to be 4.6$\times$10$^{9}$ $\nu \, cm^{-2}sec^{-1}$ on the Earth.}
\centering
\arrayrulewidth=1pt
\small
\begin{tabular}{cccccccc}
\hline\hline
\multicolumn{8}{c}{References That Conclude \textbf{Time-dependence} of Radioactive Decay Parameters}\\
\hline
 &  &  & \multirow{2}{*}{Detector}   &  \multirow{2}{*}{Measured}  &  & $\nu$ or $\overline{\nu}$  & Cross Section   \\
 Source& Reference & Mode & \multirow{2}{*}{Type}   & \multirow{2}{*}{Radiation}  & Size of Effect &  Variations & Sensitivity \\
 &  &  &   &   &  &  ($cm^{-2}sec^{-1}$) &($cm^{2}$)  \\
\hline
$^{3}$H & Falkenberg (2001)\protect\cite{Falkenberg200241} & $\beta^{-}$ & Photodiodes & $\beta^{-}$ & 3.7E-03 & 4.6E+9 & 1.4E-21 \\
$^{3}$H & Veprev(2012)\protect\cite{Veprev201226} & $\beta^{-}$ & Liq. Scintillation & $\beta^{-}$ & 2.0E-01 & 4.6E+9 & 7.8E-20 \\
$^{22}$Na/$^{44}$Ti & O'Keefe(2013)\protect\cite{OKeefe2013} & $\beta^{+}$,$\epsilon$ & Solid State (Ge) & $\gamma$ & 3.4E-04 & 4.6E+9 & 6.3E-22 \\
$^{32}$Si/$^{36}$Cl & Alburger(1986)\protect\cite{ALBURGER1986168} & $\beta^{-}$ & Gas proportional & $\beta^{-}$ & 5.0E-03 & 4.6E+9 & 1.6E-22 \\
$^{36}$Cl & Jenkins(2012)\protect\cite{Jenkins201281} & $\beta^{-}$ & Geiger-Muller & $\beta^{-}$ & 1.5E-02 & 4.6E+9 & 2.4E-25 \\
$^{54}$Mn & Jenkins(2009)\protect\cite{Jenkins2009407} & $\epsilon$ & Scintillation & $\gamma$ & 1.0E-03 & 4.6E+9 & 5.6E-21 \\
%54Mn & Jenkins(2011) & $\epsilon$ & Scintillation & $\gamma$ & 1.0E-03 & 4.6E+9 & 5.6E-21 \\
$^{60}$Co & Parkhomov (2005)\protect\cite{parkhomov2005bursts} & $\beta^{-}$ & Geiger-Muller & $\beta^{-}$,$\gamma$ & 3.0E-03 & 4.6E+9 & 2.7E-21\\
$^{60}$Co & Baurov(2007)\protect\cite{Baurov2007} & $\beta^{-}$ & Scintillation & $\gamma$ & 7.0E-03 & 4.6E+9 & 6.4E-21 \\
%60Co & Muromtsev(2012) & $\beta^{-}$ & Scintillation & $\gamma$ & 4.2E-03 & 2.0E+17$^{*}$ & 4.7E-30 \\
$^{90}$Sr/$^{90}$Y & Parkhomov (2011)\cite{Parkhomov2011} & $\beta^{-}$ & Geiger-Muller & $\beta^{-}$ & 2.3E-03 & 4.6E+9 & 3.8E-22 \\
$^{90}$Sr/$^{90}$Y & Sturrock(2012) \cite{Sturrock2012755} & $\beta^{-}$ & Geiger-Muller & $\beta^{-}$ & 2.3E-03 & 4.6E+9 & 3.8E-22 \\
$^{90}$Sr/$^{90}$Y & Sturrock(2016)\protect\cite{Sturrock20168} & $\beta^{-}$ & Liq.,Scintillation(TDCR) & $\beta^{-}$ & 2.0E-04 & 4.6E+9 & 3.3E-23 \\
$^{137}$Cs & Baurov(2007)\protect\cite{Baurov2007} & $\beta^{-}$ & Scintillation & $\gamma$ & 2.0E-03 & 4.6E+9 & 3.2E-22 \\
%198Au & Lindstrom(2011)\cite{Lindstrom2011269} & $\beta^{-}$ & Sol. St. (Ge) & $\gamma$ & 2.0E-04 & 2.3E+12$^{*}$  & 2.6E-22 \\
%137Cs & Muromtsev(2012) & $\beta^{-}$ & Scintillation & $\gamma$ & 1.6E-03 & 2.5E+17$^{*}$ & 4.6E-30 \\
$^{226}$Ra & Jenkins(2009)\protect\cite{Jenkins200942}& $\alpha$ & Ion Chamber & $\alpha$ & 2.0E-03 & 4.6E+9 & 6.0E-24 \\
\hline\hline
\end{tabular}
\label{limit_yes}
%\end{sidewaystable}
\end{table*}
%\restoregeometry

%%%%%%%%%%%%%%%%%%%%%%%%%%%%%%%%%%%%%%%%%%%%%

%\newgeometry{left=1cm,right=1cm,top=2cm}
\begin{table*}
%\begin{sidewaystable} %table rotate 90,  create a new page
%\setstretch{1}  %Adjust the thickness of line
\caption[Summary of experiments that conclude \textbf{Null} evidence for radioactive decay rate parameter variation.]{Summary of experiments that conclude \textbf{Null} evidence for radioactive decay rate parameter variation. Estimation of the interaction cross section is based on Eq.~\ref{crosssection1} by inputting the associated size of the effect, mean lifetime as well as neutrino or antineutrino flux from each reference. The variation of the solar neutrino flux is taken to be 4.6$\times$10$^{9}$ $\nu \, cm^{-2}sec^{-1}$ on the Earth. * indicates the reactor antineutrino flux at various locations.}
\arrayrulewidth=1pt
\small
\centering
\begin{tabular}{cccccccc}
\hline\hline
\multicolumn{8}{c}{References That Conclude \textbf{Null} Evidence of Radioactive Decay Rate Parameters Variation}\\
\hline
 &  &  & \multirow{2}{*}{Detector}   &  \multirow{2}{*}{Measured}  &  &  $\nu$ or $\overline{\nu}$  & Cross Section   \\
 Source& Reference & Mode & \multirow{2}{*}{Type}   & \multirow{2}{*}{Radiation}    & Sensitivity &  Variations & Sensitivity \\
 &  &  &   &   &  &  ($cm^{-2}sec^{-1}$) & ($cm^{2}$) \\
\hline
$^{3}$H & Bruhn(2002)\protect\cite{Bruhn200228} & $\beta^{-}$ & Photodiode & $\beta^{-}$ & 2.0E-03 & 4.6E+9 & 7.8E-22 \\
$^{22}$Na/$^{44}$Ti & Norman(2009)\protect\cite{Norman2009135} & $\beta^{+}$,$\epsilon$ & Solid State (Ge) & $\gamma$ & 1.8E-03 & 4.6E+9 & 3.2E-21 \\
$^{22}$Na & Meijer(2011)\protect\cite{deMeijer2011320} & $\beta^{+}$ & Solid State (Ge) & $\gamma$ & 2.0E-04 & 5.0E+10$^{*}$  & 3.4E-23 \\
$^{22}$Na & Meijer(2014)\protect\cite{upeerlimit} & $\beta^{+}$ & Solid State (Ge) & $\gamma$ & 5.1E-05 & 1.6E+13$^{*}$  & 2.7E-26 \\
$^{32}$Si/$^{36}$Cl & Semkow(2009)\protect\cite{Semkow2009415} & $\beta^{-}$ & Gas proportional & $\beta^{-}$ & 1.5E-03 & 4.6E+9 & 4.7E-23 \\
$^{36}$Cl & Kossert(2014)\protect\cite{Kossert201433} & $\beta^{-}$ & Liq. Scintillation(TDCR) & $\beta^{-}$ & 4.0E-04 & 4.6E+9 & 6.4E-27 \\
$^{40}$K & Bellotti(2013)\cite{Bellotti2013116} & $\epsilon$ & Scintillation(NaI) & $\gamma$ & 1.0E-04 & 4.6E+9 & 4.0E-32 \\
$^{54}$Mn & Meijer(2011)\protect\cite{deMeijer2011320} & $\epsilon$ & Solid State (Ge) & $\gamma$ & 4.0E-04 & 5.0E+10$^{*}$  & 2.1E-22 \\
$^{90}$Sr/$^{90}$Y & Kossert(2015)\protect\cite{Kossert201518} & $\beta^{-}$ & Liq. Scintillation(TDCR) & $\beta^{-}$ & 3.0E-04 & 4.6E+9 & 5.0E-23 \\
$^{85}$Kr & Schrader (2010)\protect\cite{Schrader20101583} & $\beta^{-}$ & Ion Chamber & $\gamma$ & 5.0E-04 & 4.6E+9 & 2.2E-22 \\
$^{108m}$Ag & Schrader (2010)\cite{Schrader20101583} & $\epsilon$ & Ion Chamber & $\gamma$ & 9.0E-03 & 4.6E+9 & 9.9E-23 \\
$^{133}$Ba & Schrader (2010)\protect\cite{Schrader20101583} & $\beta^{-}$ & Ion Chamber & $\gamma$ & 1.5E-03 & 4.6E+9 & 6.9E-22 \\
$^{137}$Cs & Bellotti(2013)\protect\cite{Bellotti2013116} & $\beta^{-}$ & Scintillation & $\gamma$ & 8.5E-05 & 4.6E+9 & 1.4E-23 \\
$^{137}$Cs & Schrader (2010)\protect\cite{Schrader20101583} & $\beta^{-}$ & Ion Chamber & $\gamma$ & 4.6E-04 & 4.6E+9 & 7.4E-23 \\
$^{137}$Cs & Meijer(2011)\protect\cite{deMeijer2011320} & $\beta^{-}$ & Solid State (Ge) & $\gamma$ & 1.7E-04 & 5.0E+10$^{*}$  & 2.5E-24 \\
$^{152}$Eu & Meijer(2011)\protect\cite{deMeijer2011320} & $\beta^{-}$,$\epsilon$ & Sol. St. (Ge) & $\gamma$ & 1.4E-04 & 5.0E+10$^{*}$  & 4.5E-24 \\
$^{152}$Eu & Siegert(1998)\protect\cite{Siegert19981397} & $\beta^{-}$,$\epsilon$ & Ion Chamber & $\gamma$ & 5.0E-04 & 4.6E+9 & 1.8E-22 \\
$^{152}$Eu & Siegert(1998)\protect\cite{Siegert19981397} & $\beta^{-}$,$\epsilon$ & Sol. St. (Ge) & $\gamma$ & 3.0E-02 & 4.6E+9 & 1.6E-21 \\
$^{152}$Eu & Schrader (2010)\protect\cite{Schrader20101583} & $\beta^{-}$,$\epsilon$ & Ion Chamber & $\gamma$ & 5.0E-04 & 4.6E+9 & 1.8E-22 \\
$^{154}$Eu & Siegert(1998)\protect\cite{Siegert19981397} & $\beta^{-}$,$\epsilon$ & Ion Chamber & $\gamma$ & 5.0E-04 & 4.6E+9 & 2.8E-22 \\
$^{154}$Eu & Siegert(1998)\protect\cite{Siegert19981397} & $\beta^{-}$,$\epsilon$ & Sol. St. (Ge) & $\gamma$ & 3.0E-02 & 4.6E+9 & 1.7E-20 \\
$^{154}$Eu & Schrader (2010)\protect\cite{Schrader20101583} & $\beta^{-}$,$\epsilon$ & Ion Chamber & $\gamma$ & 5.0E-04 & 4.6E+9 & 2.8E-22 \\
$^{155}$Eu & Siegert(1998)\protect\cite{Siegert19981397} & $\beta^{-}$ & Sol. St. (Ge) & $\gamma$ & 3.0E-02 & 4.6E+9 & 3.0E-20 \\
$^{226}$Ra & Siegert(1998)\cite{Siegert19981397} & $\alpha$ & Ion Chamber & $\alpha$ & 1.0E-03 & 4.6E+9 & 3.0E-24 \\
$^{226}$Ra & Semkow(2009)\protect\cite{Semkow2009415} & $\alpha$ & Ion Chamber & $\alpha$ & 3.0E-03 & 4.6E+9 & 9.1E-24 \\
$^{238}$Pu & Cooper(2009)\protect\cite{Cooper2009267} & $\alpha$ & Radioisotope Thermoelectric & $\alpha$ & 8.4E-05 & 4.6E+9 & 4.6E-24\\
\hline\hline
\end{tabular}
\label{limit_no}
\end{table*}
\section{Cross Section Sensitivity to Variations of Decay Rate Parameter}
\indent
To accomplish a unified frame work the induced decay rate parameter variation is framed in terms of a cross section. The standard decay rate is given by
\begin{equation}
R(t) =N_{0} \lambda \exp \left( -\lambda t \right) = \dfrac{N_{0}}{\tau } \exp \left( \dfrac{-t}{\tau } \right)
\end{equation}
where $N_{0}$ is the constant of integration which gives the original number of nuclei present when exposure begins. $\lambda$ is the decay constant, and $\tau$ is the mean lifetime which also equals to $ 1/\lambda$. The variation of the radioactive decay rate parameter at time $t$ results in
\begin{equation}
R'(t) =\frac{N_{0}}{\tau +\delta \tau} \exp\left( \dfrac{-t}{\tau + \delta \tau} \right) 
\label{rate_vary}
\end{equation}
where $\delta \tau$ is the measured variation. The experimental cross-section expresses the reaction rate to the exposure from the source, and can be written as the (Reaction Events per Unit Time per Nucleus) divided by the (Incident Flux of Neutrinos per Unit Area Per Unit Time).
\indent
This is equivalent to  
\begin{equation}
\sigma = \left\vert \dfrac{\delta R(t)}{N(t)} \right\vert \times \frac{1}{{{\Delta F_{{\nu ^{}}o{r^{}}\bar \nu }}}}=  \frac{\left\vert \delta R(t)/R(t) \right\vert}{\tau \times {{\Delta F_{{\nu ^{}}o{r^{}}\bar \nu }}}} 
\label{cross_define}
\end{equation}
where $\delta R(t)$ is the reaction events per unit time. $N(t)$ is the number of nuclei at time $t$. ${{\Delta F_{{\nu ^{}}o{r^{}}\bar \nu }}}$ is the variation of the neutrino or antineutrino flux. In addition the definition $N(t)=R(t) / \lambda$ has been used.\\
\indent
To convert $\left\vert \delta R(t)/N (t)\right\vert$ into the measuring limit $\delta \lambda/\lambda$, the reaction rate per unit time per nucleus is given by
\begin{widetext}
\begin{equation}
\begin{split}
    \frac{{{\delta R(t)}}}{N(t)} & = \dfrac{R(t)-R'(t)}{N(t)} 
    = \frac{{{N_0}\left[ {{\tau ^{ - 1}}\exp \left( { - t/\tau } \right) - {{\left( {\tau  + \delta \tau } \right)}^{ - 1}}\exp \left( { - t/(\tau  + \delta \tau )} \right)} \right]}}{{{N_o}\exp \left( { - t/\tau } \right)}} \\
    & = \frac{1}{\tau } - \frac{1}{{\left( {\tau  + \delta \tau } \right)}}\exp \left( {\frac{{ - t}}{{\tau  + \delta \tau }}} \right)\exp \left( {\frac{t}{\tau }} \right)\\
\end{split}
\label{R_367}
\end{equation}
\end{widetext}
Because $\delta \tau \ll \tau$,
\begin{equation}
\begin{split}
\exp \left( {\frac{{ - t}}{{\tau  + \delta \tau }}} \right) & \sim \exp \left( { - \frac{t}{\tau }\left( {1 - \frac{{\delta \tau }}{\tau }} \right)} \right) \\
& = \exp \left( { - \frac{t}{\tau }} \right)\exp \left( {\frac{{\delta \tau }}{\tau }\frac{t}{\tau }} \right), \\
\end{split}
\end{equation}
Using these approximations in Eq~\ref{R_367} results in
\begin{equation}
\begin{split}
    \frac{{{\delta R(t)}}}{N(t)} & = \frac{1}{\tau }\left[ {1 - {{\left( {1 + \frac{{\delta \tau }}{\tau }} \right)}^{ - 1}}\exp \left( {\frac{{\delta \tau }}{\tau }\frac{t}{\tau }} \right)} \right].\\
\end{split}
\label{R_376}
\end{equation}
Even with significant decay ($t \sim \tau$), the exponential term in Eq.~\ref{R_376} is still small because $\delta \tau /\tau $ ranges from $10^{-2}$ to $10^{-5}$. From this consideration
\begin{equation}
    \begin{split}
    \frac{{{\delta R(t)}}}{N(t)} &=\frac{1}{\tau }\left[ {1 - \left( {1 - \frac{{\delta \tau }}{\tau }} \right)\left( {1 + \frac{{\delta \tau }}{\tau }\frac{t}{\tau }} \right)} \right]\\
    & = \frac{1}{\tau }\left[ {\frac{{\delta \tau }}{\tau }\left( {1 - \frac{t}{\tau } 
    + \frac{{\delta \tau }}{\tau }\frac{t}{\tau }} \right)} \right]\\
    \end{split}
\end{equation}
In this experiment as well as all those reviewed the measurement time $t \ll \tau$, therefore,
\begin{equation}
        \frac{{{\delta R(t) }}}{N(t)} \sim \frac{1}{\tau }\frac{{\delta \tau }}{\tau }
        \label{R_380}
\end{equation}
\indent
All the reviewed results as well as this experiment can be framed as a cross section for comparison by dividing by the neutrino flux variation. Inputting the variation decay rate, Eq.~\ref{R_380} into Eq.~\ref{cross_define} yields,
\begin{equation}
\sigma = \frac{{ \left\vert \delta \tau / \tau \right\vert }}{{\tau  \times {\Delta F_{{\nu ^{}}o{r^{}}\bar \nu }}}}
\label{crosssection_L}
\end{equation}
where $\Delta F_{{\nu ^{}}o{r^{}}\bar \nu }$ is the variation of the neutrino or antineutrino flux. With $\delta \tau = -\delta \lambda / \lambda^{2}$, and $\tau=1/\lambda$, Eq.~\ref{crosssection_L} can be written
\begin{equation}
\sigma  = \frac{{\left\vert \delta \lambda / \lambda\right\vert}}{{\tau  \times {\Delta F_{{\nu ^{}}o{r^{}}\bar \nu }}}}
\label{crosssection1}
\end{equation}
\indent
Using this frame work a cross section or cross section sensitivity limit can be assigned to each experiment reported in the literature. In addition, two routes to improve the cross section sensitivity for a given isotope are presented. The first, is to measure the isotope decay parameter, ${\left\vert \delta \lambda / \lambda\right\vert}$ as precisely as possible. The second, is to get as close as possible to the reactor core to increase the on/off antineutrino flux variation, ${\Delta F_{{\nu ^{}}o{r^{}}\bar \nu }}$. This experiment has taken both approaches.

\section{Solar Neutrino Flux}
 Again, while the source of the reported rate variations in not known, if Solar in origin then it is reasonable to assume it is proportional to the Solar neutron flux. As is well known the source of solar neutrino production is the fusion reaction of hydrogen into helium resulting in 2$\nu_e$ produced by the end of the process. The result is a total solar neutrino flux on the Earth of $6.5 \times 10^{10}\,cm^{-2}sec^{-1}$, assuming no $\nu_e$ oscillations  \cite{Bellerive200419}. Also well known is the Davis result indicating $\sim1/2$ these solar neutrinos have oscillated before reaching the Earth\cite{Davis196820}. However, for the purpose of comparison, the total solar neutrino flux on the Earth in this paper is taken to be $6.5 \times 10^{10}\, \nu \,cm^{-2}s^{-1}$.\\
\indent
As shown, it is the variation in the solar flux that causes the decay rate parameter variation not the flux. Due to the Earths elliptical orbit, with perihelion at 147.1 Mkm, aphelion at 152.1 Mkm and having a semi-major axis of 149.6 Mkm, the solar neutrino flux variation is $\sim$7\% found by, 
\begin{equation}
\begin{split}
\dfrac{r_{perihelion}^{-2} - r_{aphelion}^{-2}}{r_{semi-major}^{-2}} & =\dfrac{(147.1)^{-2}- (152.1)^{-2}}{(149.6)^{-2}}\\
& \\
\end{split}
\end{equation} 
Thus, the amplitude of the variation of the solar neutrino flux is $\sim 4.6\times10^{9}\,\nu\,cm^{-2}sec^{-1}$ on the Earth.\\
\indent
With this framework, the results in the literature can be compared and are displayed in Table~\ref{limit_yes} and Table~\ref{limit_no}.  

\section{The High Flux Isotope Reactor} 
\begin{figure}[ht!]
\begin{center}
\includegraphics[scale=.41]{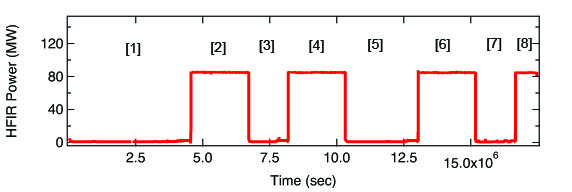} 
\end{center}
\caption{The HFIR power as a function of time when the $^{54}$Mn source was present in the experiment. Each Period of the experiment is indicated.}
\label{HFIR_Power}
\end{figure}

The antineutrino source for this experiment is the High Flux Isotope Reactor (HFIR) located at Oak Ridge National Laboratory at Tennessee, USA. HFIR uses highly enriched $^{235}$U (HEU) as the fuel. The operating cycle  consists of full-power operation for approximately 23-27 days using a newly constructed core for each cycle. Figure~\ref{HFIR_Power} shows the HFIR reactor power as a function of time during the $^{54}$Mn phase of the experiment. The average operating power is calculated from the recorded reactor power data taken every second, and is 86.007$\pm$0.22 MW. The reactor power is very stable with a variance of $\delta p/p_{mean} \sim 2.6 \times 10^{-3}$,
where $\delta p$ is the standard deviation of the average power, and $p_{mean}$ is the average power of HFIR in operation.\\
\indent
The average antineutrino flux can be estimated knowing the fissile fuel composition, highly enriched $^{235}$U, and the reactor's thermal power. 
The antineutrino production in the core is
\begin{equation}
\dfrac{dN_{\overline{\nu}}}{dt}=n_{\overline{\nu}} \times \dfrac{P_{H}}{E_{F}}
\end{equation}
where $P_{H}$ is the average thermal power output, $E_{F}$ is average released thermal energy per fission, and $n_{\overline{\nu}}$ is the average number of antineutrino generated per fission.\\ 
\indent
Table~\ref{type_of_fuel} displays the parameters used to find the antineutrino flux for this experiment assuming all the production is from $^{235}$U. The thermal energy released per each fission of $^{235}$U is
\begin{equation}
\begin{split}
E_{f}&=E_{F} - \langle E_{\overline{\nu}}\rangle \times n_{\overline{\nu}}\\
     &=201.7 - 1.46 \times 5.58 \\
     & = 193.6\,MeV \\
\end{split}
\end{equation}
where $E_{F}$ is the released energy per fission, and $\langle E_{\overline{\nu}}\rangle$ is the mean energy of the antineutrinos. Hence, the estimated rate, $\dot{N}_{\overline{\nu}}$ from the HFIR reactor core is given by 
\begin{equation}
%\begin{split}
    \dot{N}_{\overline{\nu}} = 5.58 \times \dfrac{86.007\,MW}{193.6\,MeV} 
   = 1.53 \times 10^{19}\,\overline{\nu}\,sec^{-1}.
%\end{split}
\end{equation}
The antineutrino flux at the HPGe detector face, located $6.53\,m$ from the core is estimated to be $2.86 \times 10^{12}\, \overline{\nu} \,cm^{-2}sec^{-1}$, assuming the core is a point source. This flux is nearly $50$ times higher than the solar neutrino flux on the Earth, and the reactors on/off cycles produce more than 600 times larger variation than the solar neutrino flux variation.
\begin{table}[ht!]
\caption {Characteristics of Antineutrino production from $^{235}$U\protect\cite{Kessler}.}
\begin{center}   
%\setstretch{1.25}
\begin{tabular}{cc}
        \hline\hline 
    Type of Fuel & $^{235}$U \\
        \hline 
	Released energy per fission ($E_{F}$)(MeV) & 201.7 \\
	Mean energy of $\overline{\nu}$ ($\langle E_{\overline{\nu}}\rangle$)(MeV) & 1.46 \\
	Number of $\overline{\nu}$ per fission ($n_{\overline{\nu}}$)($E>1.8$ MeV) & 5.58 \\
		 \hline\hline        
\end{tabular}       
\end{center}
\label{type_of_fuel}
\end{table}

\section{Experimental Configurations}
\begin{figure}[t!]
\begin{center}
\includegraphics[scale=.8]{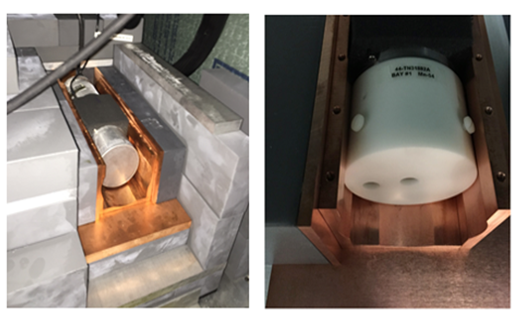} 
\end{center}
\caption[HPGe Configuration.]{HPGe Configuration: (Left) positioned in the copper florescence shielding without the Source Cup. (Right) Source Cup in place.  Button source to be place in the center cavity.}
\label{HPGe1}
\end{figure}

The experiment used a 60$\%$, N-type, High Purity Germanium Detector Spectrometer (HPGe) system, employing first $^{54}$Mn, then $^{137}_{~55}$Cs in the same experimental configuration. The $\sim$1$\mu$Ci button sources were held fixed by a polycoarbonate cup, on the detectors central axis about 2 cm from the detectors face as shown in Figure~\ref{HPGe1}. The starting count rate was $\sim$8kcps.  The HPGe energy resolution (FWHM) $\delta E/E \sim 1.67\times10^{-3}$ was measured at $E_{\gamma}=1.33$ MeV using a $^{60}$Co source.\\
\indent
The shielding house\cite{heim}, consisted of a 1 inch boronated-polyethylene (b-poly) skin, followed by 4 inches of Pb and an inner liner of 1 inch b-poly. The b-poly served as a neutron shield. The detectors Ge-crystal vacuum housing consisted of an aluminum can with a Be-window.  Surrounding the vacuum housing was a 5/8 inch thick copper box to absorb the $\sim$100 keV Pb fluorescence photons. Additional thin plates of aluminum shielding were placed around the source holder cup to improve the absorption of the $\sim$10 keV Cu fluorescence photons. Outside the box was an additional 4 inches of Pb shielding for a total of at least 8 inches of Pb shielding surrounding the detector having a total wt. of 5.5 tons. The inner bulk structure is contained within a polystyrene enclosure, whose temperature is maintained by a PID-controlled thermoelectric unit (TECA AHP-1200HC). A set point of 10 $^o$C was chosen for the enclosure based on its observed nominal equilibrium temperature of 12 $^o$C - due to cooling by the detectors only - as well as the advantages of operating the thermoelectric unit strictly in one mode when possible. The enclosure is also continuously purged with nitrogen gas to prevent condensation, which can degrade spectral resolution, and to mitigate influx of contaminants such as $^{226}$Ra and $^{41}$Ar. System power is routed through uninterruptible power supplies (UPS), which provide conditioning as well as approximately 30 minutes of backup power in the event of power loss or necessary platform relocation. \\
\indent
It must be noted that while the detector and source were protected within to the highly controlled shielding house and its environment, the refrigerator, an X-Cooler III and the electronic spectrometer were mounted on the roof of the shielding house and thus exposed to the reactor buildings environmental systems.

\section{Shielding Performance and Background Stability}
\begin{figure}[t!]
\begin{center}
\includegraphics[scale=.4]{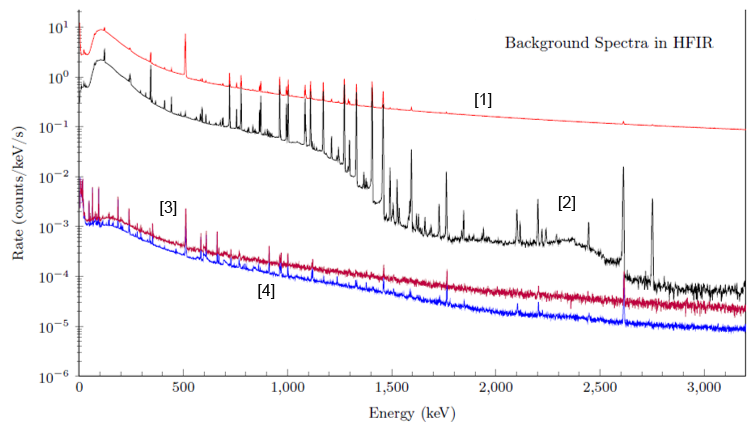}
\end{center}
\caption[HFIR background spectra, ]{HFIR background spectra, [1] unshielded reactor-on, [2] unshielded reactor-off, [3] fully shielded reactor-on, and [4] fully shielded reactor-off.}
\label{BG_Compare}
\end{figure}

\begin{figure}[t!]
\begin{center}
\includegraphics[scale=.4]{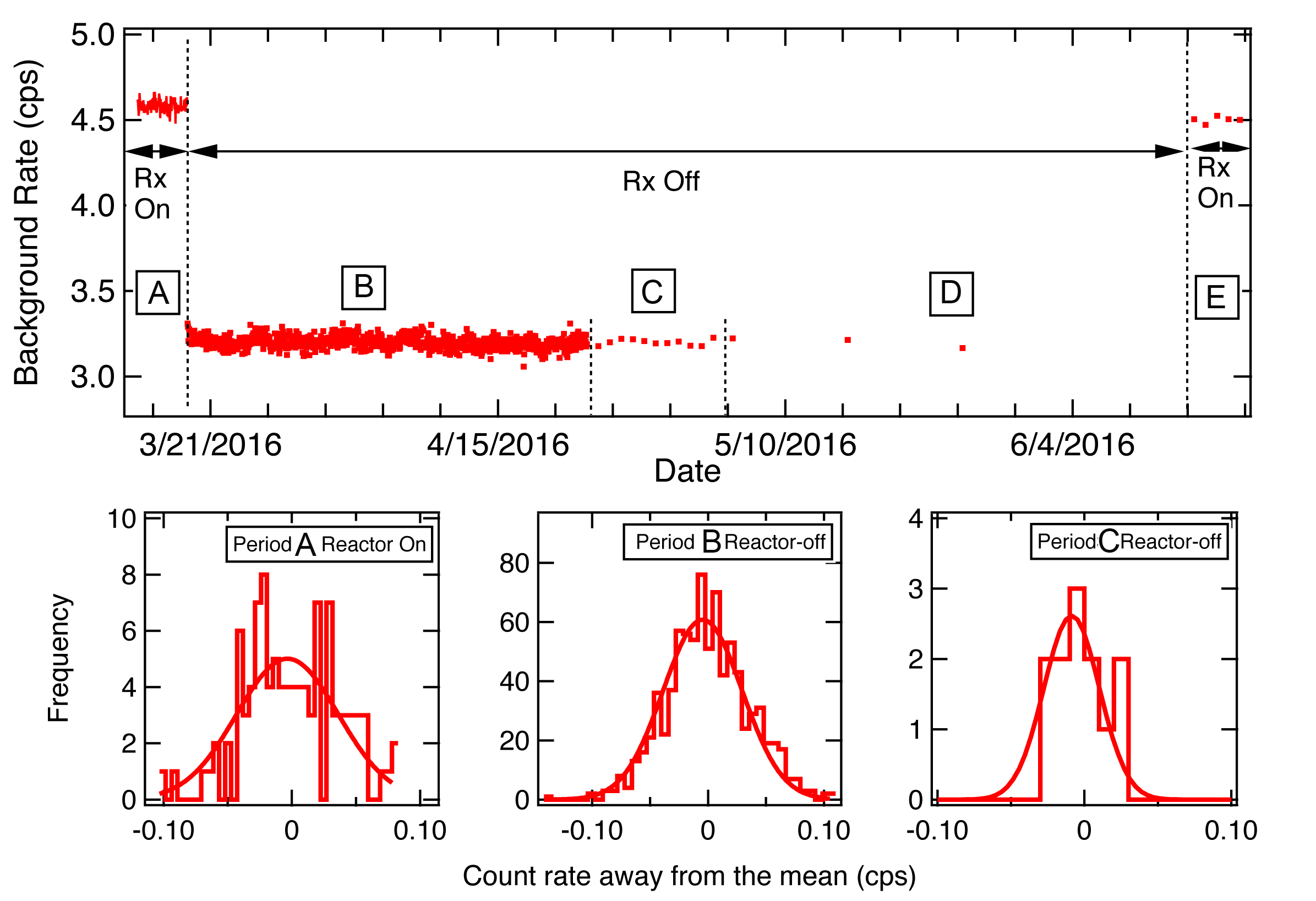} 
\end{center}
\caption[Full spectrum background rate as a function of time including reactor-on, and reactor-off cycles.]{(Top) Full spectrum background rate as a function of time including reactor-on, and reactor-off cycles.  The source is not present. (Left Lower) The distribution away from the mean in count rate for 1-hour runs during the Period A, reactor-on. (Center Lower) The distribution away from the mean in count rate for 1-hour runs during Period B, reactor-off. (Right Lower) The distribution away from the mean in count rate for 1-day runs during Period C, reactor-off.}
\label{Bg_Rate_onoff}
\end{figure}

The background radiation at the experimental location in HFIR is due to (1) the neutrons, and gamma rays directly produced from the reactor operation, (2) neutron activated building components, (3) scattered radiation from nearby beamline operations, (4) decay radiation from a nearby source storage room, (5) the shielding materials, (6) natural radioactivity within the building, (7) trace contamination due to the presence of special nuclear material in the building, (8) cosmic radiation, and finally (9) the HPGe detector itself.  Because the detector is only 6.53 meter from the reactor core gamma rays and neutrons produced during reactor on and off periods are significant if the detector were left unshielded.  Figure~\ref{BG_Compare} shows the background radiation with, and without shielding during reactor-on and off periods. The spectra show that the shielding is effective in suppressing the backgrounds by a factor greater than $1.8 \times 10^{-4}$ during reactor-on periods, and $6.7 \times 10^{-4}$  during reactor-off periods.\\
\indent
The reactor on and off background spectra was assumed to be stable over the course of the experiment.  To check the time-dependence stability of the background rate, data were collected for nearly 90 days consisting of hourly, daily, and 10 day background runs starting 03/09/2016.  The background spectra were collected in the shielding house, which included two reactor-on periods, and one reactor-off period.  Figure~\ref{Bg_Rate_onoff} shows the full background spectrum rate as a function of time. Table~\ref{Bg_Stability} summarizes the information during each background collection period. The timing error contribution from the DSPEC-50 are estimated to be below $8.13\times10^{-7}$cps which is negligible but tracked throughout the analysis. \\
\indent
The rate distributions for the runs labeled as Period A, Period B, and Period C are Gaussian as shown in Figure~\ref{Bg_Rate_onoff}. The width of the distributions are in agreement with the estimate found using the averaged error of the mean found for each run. As shown in Table~\ref{Bg_Stability}, the statistical mean run error, and the width of the run rate distribution are in agreement. This means the calculated errors are correct as proven by the distributions.\\
\indent
The means of three reactor-off Periods B, C, and D are in excellent agreement with their individual variation from the averaged mean shown in the standard deviation row in Table~\ref{Bg_Stability}. These results prove the background is exceptionally stable over the 74-days reactor-off period. All the reactor-off data are used for the background subtraction. The two reactor-on Periods, A and E are in disagreement at the level of $0.04\pm0.005$cps from the mean. While the disagreement is small, it is significant. The Period A data was discarded. Just before the background runs were taken the shielding was deconstructed, followed by a week long period of reconstruction with Period A starting immediately afterwards. This work exposed the shielding to a different and uncontrolled temperature environment. For this reason the first 5-days of data, shown as Period A in Figure \ref{Bg_Rate_onoff}, is considered unstable as the detector and shielding system needed time to reach equilibrium. Thus the reactor-on background used for corrections only included Period E, consisting of the average of 5, 1-day runs.\\
\indent
The average full spectrum rate for the reactor-on background is $4.502\pm0.003$cps from Period E. The average rate for the reactor-off background is $3.2003\pm0.0007$cps using Period B, Period C, and Period D, that is the average of the 74 days of runs.\\
\indent The full spectrum average rate of $^{54}$Mn is $\sim 6500$cps and $^{137}$Cs is $\sim 8300$cps. From this, the estimated contribution of the reactor-on background spectrum to the overall error is less than $\sim 6\times10^{-7}$, and the reactor-off background contribution is less than $\sim1\times10^{-7}$. All are small enough to be neglected but are tracked throughout the analysis.
%\begin{sidewaystable}[ht!]
\begin{table*}[ht!]
%\setstretch{1.25}
\caption[The average background rate, and associated error during the reactor-on or reactor-off periods at HFIR.]{The average background rate, and associated error during the reactor-on or reactor-off periods at HFIR. The distribution error is not available in Period D, and E due to lack of data points. S.D. means the standard deviation. The S.D from the mean is not available for reactor-on Period A, and E because Period A data
was dropped from the analysis (see text)}
\centering
\begin{tabular}{cccccc}\hline\hline
&  Period A  &  Period B & Period C & Period D & Period E \\
\hline
Reactor status & on & off & off & off & on \\ 
Duration (Days) & 5 & 32 & 12 & 30 & 5  \\
Single run time & 1-hour & 1-hour & 1-day & 10-day & 1-day \\
Timing error(cps) & 8.13E-07 & 6.77E-07 & 2.65E-08 & 2.66E-09 & 3.00E-08\\
Statistical mean run error(cps) & 3.57E-02 & 2.99E-02 & 6.09E-03 & 1.93E-03 & 7.32E-03\\ 
Distribution error(cps) & 3.95E-02 & 3.34E-02 & 1.80E-2 & N/A & N/A\\ 
Mean rate(cps) & 4.581$\pm$0.004 & 3.2001$\pm$0.001 & 3.199$\pm$0.002 & 3.201$\pm$0.001 & 4.501$\pm$0.003 \\
Reactor-off S.D. from the mean & N/A & 0.28  & 0.63 & 0.74 & N/A \\ \hline
\multicolumn{3}{r}{The average rate in all reactor-off periods :} & \multicolumn{2}{l}{$3.2003\pm0.0007$ cps} \\
\multicolumn{3}{r}{The average rate in all reactor-on periods :} & \multicolumn{2}{l}{ $4.502\pm0.003$ cps} \\
\hline\hline
\end{tabular}
\label{Bg_Stability}
%\end{sidewaystable}
\end{table*}
\section{Data Collection}
The $^{54}$Mn $\gamma$ spectra were collected from August 14, 2015, at 23:58:22 to March 09, 2016 at 23:12:51 which includes four reactor-off, and four reactor-on periods. The total running period consisted of 95 reactor-on days, and 114 reactor-off days totaling over 209 days of data collection.  The initial rate was $8.01$ kcps with a dead time of $\sim 13\% $. The ending data rate was $5.05$ kcps with a dead time of $\sim 8\%$. \\
\indent
The starting spectrum is shown in Figure~\ref{mn_FullSpectrum}.  The repetition of the photopeak or pileup peaks are observed with their relative strengths. The strength of the 3rd pileup photopeak relative to the first or true photopeak is $\sim10^{-6}$.\\
\indent
Similarly, the $^{137}$Cs gamma spectra were collected from July 03, 2016 at 13:07:37 to November 12, 2016 at  12:16:08 which includes 3 reactor-off, and 3 reactor-on periods. The total running period consisted of  49 reactor-on days and  84 reactor-off days over 133 days of data collection. The initial rate was 8.30 kcps with a dead time of 13.2\%. The ending data rate was 8.23 kcps with a dead time of  13.0\%.\\
\indent
\noindent
For both sources a single spectrum consisted of 24 hours of data collection. The computer clock was continuously synced with stratum 1 public NTP servers via the Meinberg  Time Server Monitor software. The software makes corrections to the PC clock gradually, as opposed to immediate 'step' corrections. As a check the clock offset between August 14, 2015 and April 10, 2015 exhibited a mean value of 0.14 ms and standard deviation of 21 ms. Thus the errors due to timing are negligible, at the level of $\sim$3 x 10$^{-7}$. 
\section{Data Analysis}
The goal of the analysis was to achieve a sensitivity at the level of 10$^{-5}$, by keeping the systematic errors small, allowing the statistical error to dominate the measurements.  To accomplish this the analysis proceeded as follows;  (1) First, each spectrum was independently energy calibrated.  (2) Using the dead time as reported by the electronic spectrometer, each spectrum is dead time corrected.  (3) The background spectra are energy matched to the source spectra to account for the difference in calibrations. The background spectrum is next time scaled and amplitude matched; found by comparison to a high energy region of the source spectrum. Finally the background is subtracted from the targeted source spectrum. It should be noted that the source spectrum binning width is not adjusted in this process. (4) A de-convolution algorithm is applied in order to find the pile-up free source spectrum. (5) The energy region of interest is defined which forms the signal to be tested for decay rate variations.  The region includes the photopeak and other features in order to obtain a semi-stable signal. (6) Taking advantage of the high frequency reactor on-off cycles in comparison to the approximate yearly cycle of environmental parameters, a fixed year long frequency oscillation correction is made, caused by temperature effects on the HPGe refrigerator. (7) Finally a side band correction is made to correct for nonlinear effects in the energy calibration caused by temperature and humidity effects on the electronic spectrometer housed outside the shielding housing.
\section{Energy Calibration}
To measure the calibration accurately, spectral lines are measured over the full energy range. For background spectra, 13 lines between 3 keV, and 3 MeV were used in the calibration. This assured that the energy region above the 3rd occurrence of the photopeak was well calibrated for the subtraction process.  The lines chosen are produced by neutron capture, neutron-induced ambient background, beta decay transitions induced by neutron capture, atomic fluorescence, and natural environmental backgrounds. Because both sources overwhelm these background calibration lines, the source spectra have been calibrated using (1) the indium X-ray, K-edge peak cause by the source radiation on the indium in the vacuum housing of the HPGe detector, (2) differentiation of the Compton edge, and (3) differentiation of the back-scattered peak, (4) the sources photopeak, and (5) the $^{208}$Tl gamma line from natural background.  While there are many re-occurrences of these lines as pile-up, their exact energy values were poorly understand, so could not be used for calibration.   \\
\indent
Both source and background spectra calibrations were accomplished using a two-step  process. First, a Gaussian line shape fit was used to find each centroid. Second, the energy calibration was found using a non-linear 3 parameter polynomial fit, weighted by each centroids fitting error. Due to the high statistics in each spectrum, $\sim10^9$ events, non-linear fits were required. The resulting $\chi^2$/DoF was improved by a factor of 10 over linear calibration fits.\\
\indent
The exact location of the Compton edge and back-scattered peaks are related to the detector, shielding, and source geometry and could not be found by first principles.  For this reason a single parameter iterative calibration process was performed on the first days spectrum of each source, based on minimizing the 3$^{rd}$ order polynomial fit.  Once the energy values were found, all 5 lines energy values were fixed and used to calibrate all subsequent source spectra.\\

\begin{figure}[t!]
\begin{center}
\includegraphics[scale=.45]{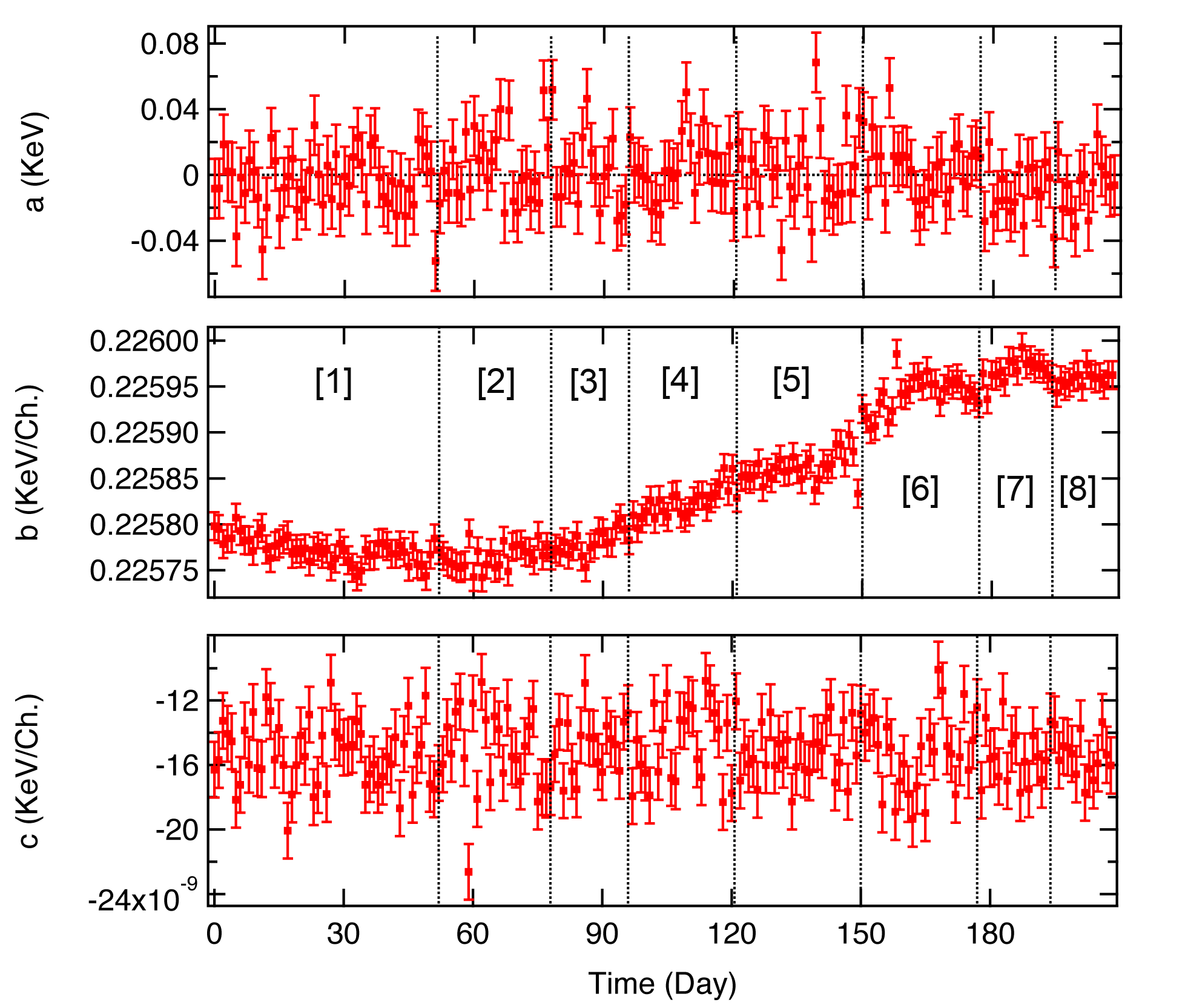}
\end{center}
\caption{Nonlinear energy calibration parameters; (Top) $a_{r}$, (Middle) $b_{r}$, and (Lower) $c_{r}$, determined for each of the daily $^{54}$Mn spectra. Each period of the experiment is indicated.}
\label{Nonlinear_5lines_abc}
\end{figure}
\begin{figure}[t!]
\begin{center}
\includegraphics[scale=.45]{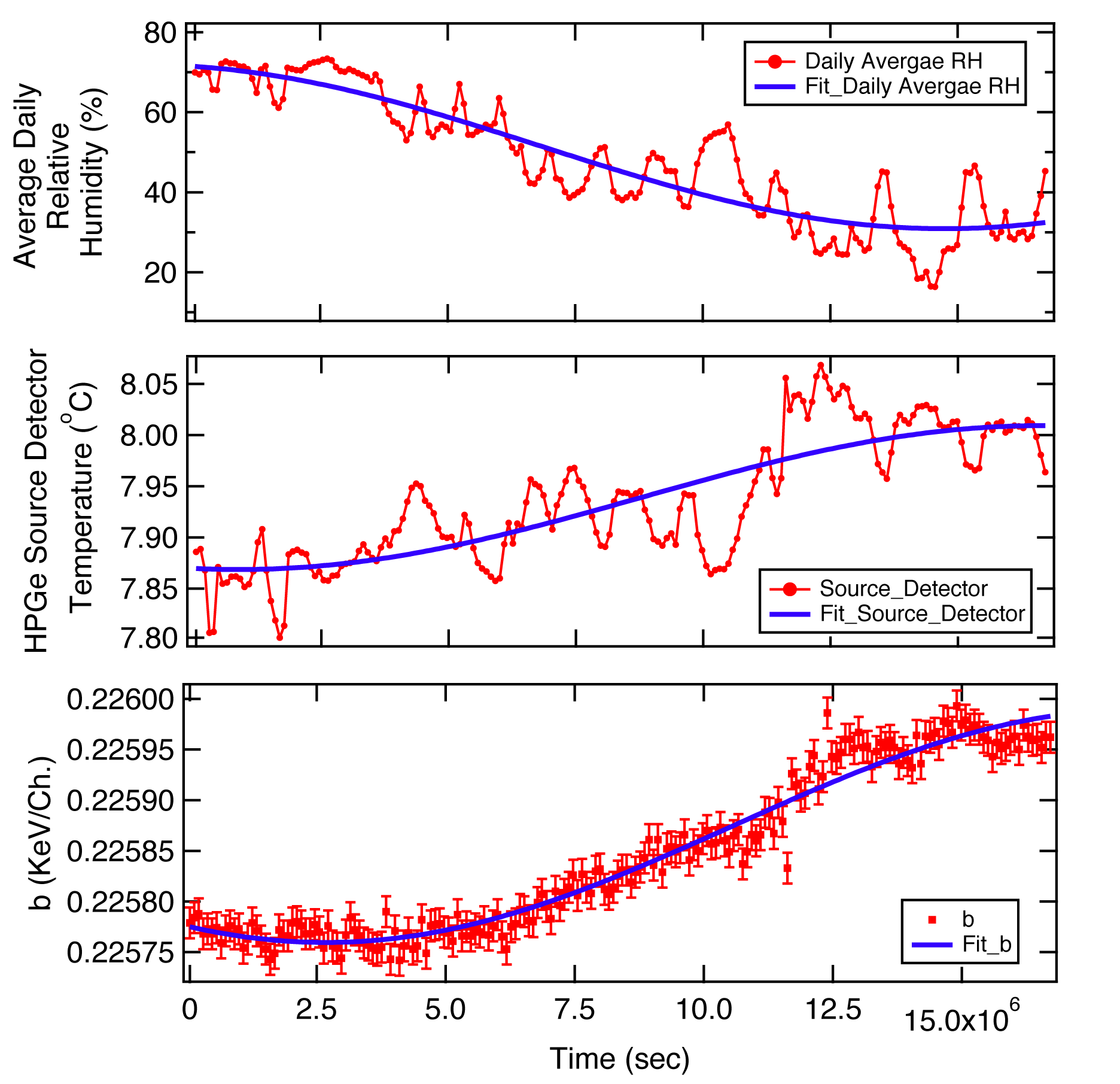} 
\end{center}
\caption{Daily averages for the humidity, temperature of the HPGe source detector, and the linear term $b_{r}$. Each is fitted with a periodic function fixed at 1 year, as a function of time.}
\label{Humindity_Temp_b}
\end{figure}

\begin{table}[t!]
\caption{The yearly cycling of the humidity, temperature for the HPGe source detector, and linear term $b$ values relative to the experiments starting date.}
\begin{center} 
\begin{tabular}{c c}
        \hline\hline Fixed $\omega$=1 year &  Negative phase ($\phi$) \\ 
        \hline     
        Out of phase humidity &   103 $\pm$ 1 day \\
         Source detector temperature &   101 $\pm$ 6 day \\
        Linear term $b$  &   122 $\pm$ 2 day \\  
        \hline\hline 
 \end{tabular}
\end{center}
\label{humidity_temp_table}
\end{table}
\section{Calibration Coefficients}
The calibration function is given in Eq. \ref{eq:cal_param}.  The fitting parameters and their fitting errors as a function of time for the  $^{54}$Mn runs are displayed in Figure \ref{Nonlinear_5lines_abc}.
\begin{equation}
E = a_{r} + b_{r}x + c_{r}x{^2}
\label{eq:cal_param}
\end{equation}
where $a_{r}$, $b_{r}$, and $c_{r}$ are the calibration coefficient of $r$th daily spectrum, and $x$ is the channel number of the spectrum.  Each parameter is discussed in turn.
\subsection{Constant Terms a$_r$}The constant terms $a_{r}$ vary about zero with variance of $\sim 0.02$ keV or less than $10\%$ of a single bin width at $0.225$ keV. The $a_{r}$ parameters measure offsets in the energy scales. $a_{r}$ is expected to be nearly zero in a low noise environment which is the case in this experiment. The stability of $a_{r}$ yields a stable Region of Interest (ROI) containing the photopeak.\\
\\
\subsection{Linear Terms b$_r$}
The linear parameter, b$_r$ is well measured with an accuracy of $\delta b/b \sim 7\times10^{-5}$. Nonetheless, the $b_{r}$ vary over the course of the experiment due to temperature variations driven by the X-cooler, the HPGe refrigerator, responding to variation in the humidity. To show this, the linear $b_{r}$ parameters are fit to a periodic function,
\begin{equation}
    b(t)=b_{0} + A sin(\omega t +\phi)
    \label{b_EQ}
\end{equation}
in which $\omega$ is fixed at 1-year. The resulting excellent fit, $\chi^{2}/Dof = 1.14$, is displayed in Figure~\ref{Humindity_Temp_b}, and the phase given in Table~\ref{humidity_temp_table}.\\
\indent
Likewise, the detector temperature is fit as a function of time, with the fit shown in Figure~\ref{Humindity_Temp_b}, and the resulting phase given in  Table~\ref{humidity_temp_table}. It should be noted that  the magnitude of $b_{r}$ is directly related to the detectors band gap which has a temperature dependence given by Varshni's empirical form\cite{VARSHNI1967149}. Varshni's form predicts that lower temperatures produce larger band gaps, and vice versa. Since lower $b_{r}$ values represent larger band gaps, b and the detector temperature should be in phase. As shown in Table~\ref{humidity_temp_table}, the phases are in reasonable agreement.\\
\indent
While the temperature of the housing enclosure is held at a very stable temperature $10.00\pm0.027\,^{o}$C. Nonetheless, the detectors X-cooler is outside the housing. The X-cooler is affected by the humidity in the following manner, as the humidity increases the heat capacity of the air increases, allowing more efficient cooling, lowering the temperature of the detector, and thus increasing the band gap. Likewise, as the humidity decreases, the heat capacity of the air decreases, yielding less efficient cooling, and allowing the detector temperature to increase, narrowing the band gap. In this way, the humidity variation should be out of phase with both the detector temperature and the $b_{r}$ parameters, as is the case shown in Figure~\ref{Humindity_Temp_b}, and Table~\ref{humidity_temp_table}, using the same fixed fit. While these effects are small, they produce yearly oscillations in the data at the level of $10^{-3}$.\\
\\
\subsection{Nonlinear Terms $c_{r}$}
The $c_{r}$ parameter is the nonlinear term, significant only for high energy spectral lines. It is due to temperature variations of the DSPEC-50 electronic spectrometer placement outside the temperature controlled housing. While in the plot $c_{r}$ appears stable, and small, $\sim10^{-8}$ keV/bin, at large channel number, for example, $\sim 4000$ where the $^{54}$Mn photopeak is located, it accounts for a shift of 1 full bin ($0.225$ keV) in the spectrum.\\
\indent
Finally, these calibration parameters result in a highly stable photopeak energy as a function of the time. The average energy of the $^{54}$Mn photopeak is $834.849 \pm 0.001$ keV, within the uncertainty of the $^{54}$Mn standard error value $\pm0.003$ keV\cite{NNDC_54MN}. Likewise the $^{137}$Cs photopeak is highly stable with a mean energy of $661.650\pm 0.011$ keV.
\section{Correction Procedures}
Because the analysis of the HPGe spectrum requires mathematical manipulation of each channel, statistical and systematic error must be considered channel by channel. There are three significant systematic errors associated with the measurement. (1) The Electronic Dead Time from the DSPEC-50 electronic spectrometer (2) Neutron, and natural background radiation which produces an unwanted background spectrum and (3) The electronic pile-up due to the inability of the electronics spectrometer to distinguish two or more pulses occurring within a time window smaller than the electronics resolving time.\\
\indent
Each correction will be discussed in order taken.  The energy calibration of the background spectra (already discussed), the dead-time produced by the digital spectrometer, background spectra re-binning and re-scaling and subtraction, the pile-up correction using a de-convolution algorithm, integration of the ROI for each 24-hour data run, a side band low frequency oscillation correction, and finally the electronics instability correction due to ambient environmental effects. 
\subsection{Dead-Time Correction}
Dead time is the integration of these periods in which the electronics do not give a response to detector pulses. The electronic spectrometer follows the non-paralyzable model.  The DSPEC-50 has two different ways it processes the 2nd incoming pulse from the HPGe detector if the previous pulse has not returned to pole-zero baseline. First, DSPEC-50 will reject the 2nd pulse if it is recognized. Second, if the 2nd pulse arrives unrecognized, that is within a shorter time than the electronic resolving time, the DSPEC-50 records the sum of two pules which is called electronic pile-up. The dead time correction would account for the total number of counts in the full spectrum if there were no electronic pile-up effects.\\ 
\indent
There are two types of pile-up effects, low energy, and high energy pile-up, defined with respect to the ROI region. The DSPEC-50 dead time correction is only correct when applied to the photopeak counts. However, the analysis uses an ROI not limited to the photopeak. The dead time correction does correctly account for the high energy pile-up, that is the loss of events out of the ROI region. However, the dead time correction does not take into account the lower energy pile-up effect. That is low energy events combining to an energy that places the combined event into the photopeak ROI. Triple pile-up is also observed at the level of $10^{-6}$ of the photopeak rate, and is neglected in this analysis. \\
\indent
Table~\ref{deadtime_table} gives the properties of the $^{54}$Mn spectrum using the corrections.
\begin{table}[t!]
\caption[Proprieties of the full $^{54}$Mn spectrum.]{Proprieties of the full $^{54}$Mn spectrum. The 1st day spectrum is in a reactor-off period. The last day of the spectrum is in a reactor-on period.}
\begin{center}   
%\setstretch{1.25}
\begin{tabular}[b]{ccc}
        \hline\hline Full Spectrum  &  First Day &  Last Day  \\
                 Daily Ratio &  (Reactor-off) &  (Reactor-on)\\
        \hline 
        	   Spectrum rate (cps) &8009.5 & 5054.5 \\  
        	   Background(Fraction)(cps) & 3.1($0.04\%$)  & 4.7 ($0.09\%$) \\  
        	   Pile-up(Fraction)(cps) &33.5 ($0.4\%$) & 13.7 ($0.04\%$)  \\  
        \hline	   
			   Dead time/day(sec) &11166.4 & 7235.9 \\        
        	   Pulse generated dead time($\mu$sec) & 16 & 16 \\
          	   Double pulse resolution($\mu$sec) & 0.5 & 0.5   \\    
         \hline 	   
          	    Background-source pileup(cps)& 0.013  & 0.012  \\
          	  
        \hline\hline        
        \end{tabular}       
\end{center}
\label{deadtime_table}
\end{table}
While it may appear optimal to finalize the dead time corrections before moving on to other corrections, this is not possible. The pile-up correction is entangled with other corrections as will be shown. The dead time correction is accurate to $10^{-3}$ if pile-up is not taken into account. The pile-up correction can be untangled by first noting in Table~\ref{deadtime_table} that the full spectrum background pile-up with the source is only 0.013 cps, calculated using the double pulse time resolution $0.5$ $\mu$sec. This rate is much smaller in the Region of Interest. For this reason, the background-source pile-up is negligible allowing the background to be first subtracted from the dead time corrected source spectrum.

\subsection{Background Re-binnig, Re-scaling, and Subtraction}
Background subtraction takes into account that the source spectra and background spectra have been calibrated using different nonlinear calibrations. As with this, and other corrections, the source spectrum is never energy scale altered during corrections. Corrections are always energy matched to the source calibration. In this way, the correction contributes the minimum possible error to the analysis. For this reason, the background correction is a three-step process. (1) The dead time correction is made channel by channel. (2) Then the background data is energy rescaled to match the difference in energy calibration parameters between the source data, and the background data. (3) Finally, the collected background is normalized to the daily rate above the third recursion of the photopeak due to pile-up. This region is unaffected by  pile-up. For the $^{54}$Mn source this range extends from $E_{low}=2515 $keV to $E_{high}$ = $3680$ keV. The correction is made using a single scaling parameter by matching the total number of counts in the source spectrum and background histograms in this comparison region. The completion of these correction are intricate and are discussed elsewhere\cite{Liu}.  After completing all the correction to the background, the resulting background spectrum is subtracted from the source spectrum.\\
\begin{figure}[ht!]
\begin{center}
\includegraphics[scale=.4]{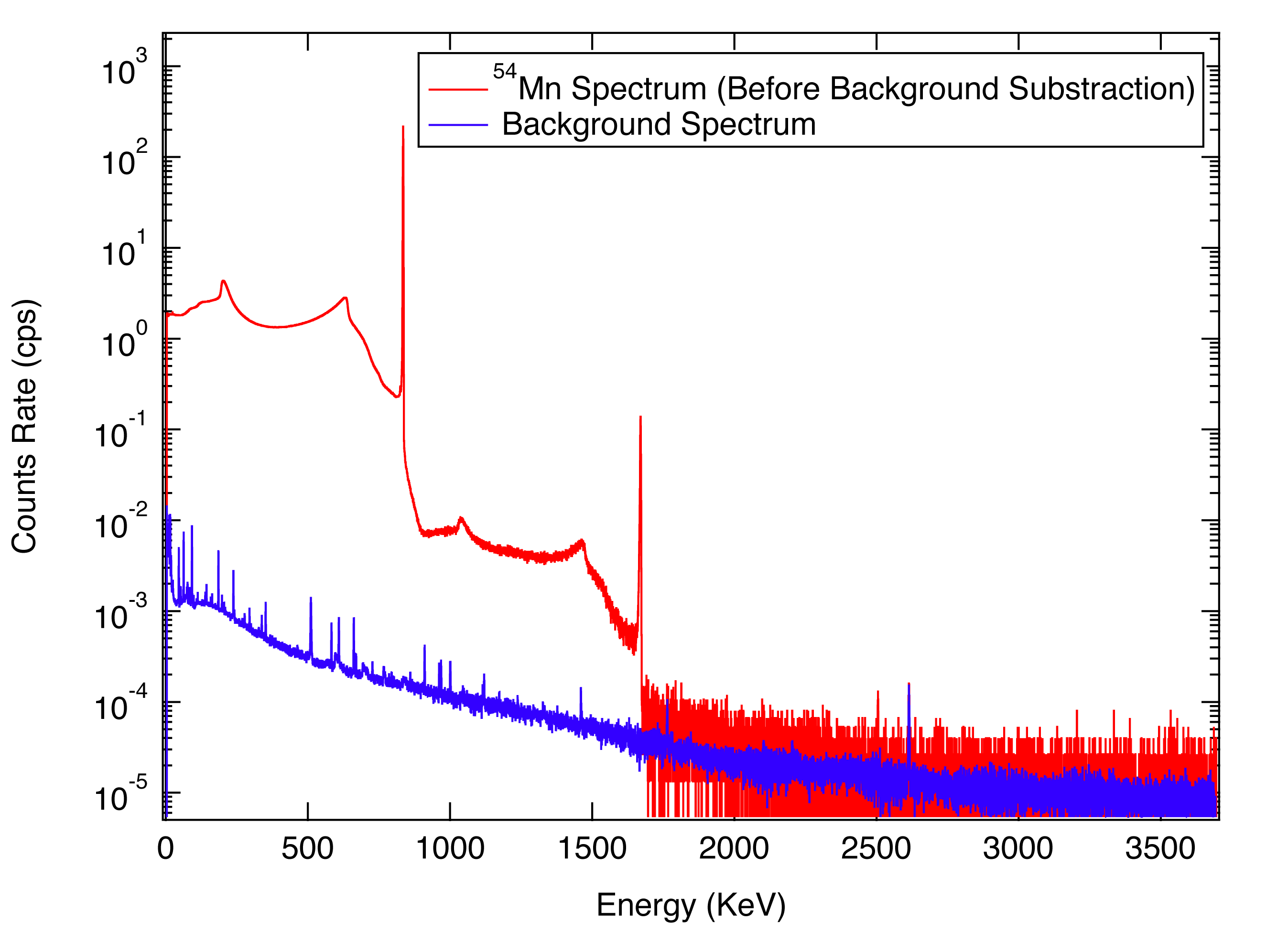} 
\end{center}
\caption[Logarithm scale of the $^{54}$Mn $\gamma$-spectrum before background spectrum subtraction, and the background spectrum]{Logarithm scale of the $^{54}$Mn $\gamma$-spectrum before background spectrum subtraction, and the background spectrum.}
\label{BG_54MnCompare}
\end{figure}
 Figure~\ref{BG_54MnCompare} compares the 1st day of the $^{54}$Mn spectrum before background correction, and the background spectrum after the matching correction. The process is channel by channel (bin by bin) subtraction using either the  reactor-on background or reactor-off background spectra. The ratio of the $^{54}$Mn spectrum's error to the error caused by amplitude scaling of the background is
\begin{equation}
        \dfrac{\sigma(i)_{amplitude}}{\sigma(i)_{source}} =
    \left\{   
    \begin{array}{ll}
    2 \times 10^{-3}\; \mathrm{\:Compton\:Region}\\
    1 \times 10^{-4}\; \mathrm{\:ROI\:Region} \\
    \end{array}
    \right\}
\end{equation}
The background statistical and re-binning error $\sigma(i)_{background}$, and the amplitude scaling error $\sigma(i)_{amplitude}$ are extremely small compared to the statistical error in the $^{54}$Mn spectrum, and are dropped. The corrected spectrum uncertainty $\sigma(i)_{corrected}$  in the $i$th bin is 
\begin{equation}
    \sigma(i)_{corrected} \sim \sigma(i)_{source} 
\end{equation}
Thus, the background subtraction contributes no error to the $^{54}$Mn spectrum.

\subsection{Pile-up Correction by De-convolution Algorithm}
\begin{figure}[t!]
\begin{center}
\includegraphics[scale=.36]{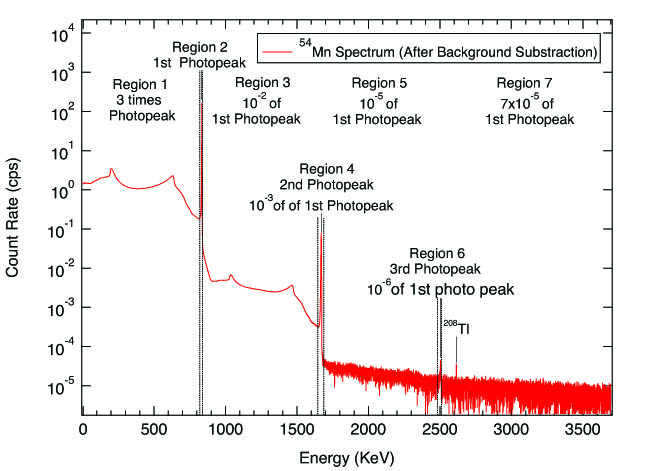} 
\end{center}
\caption[Logarithm scale of the $^{54}$Mn $\gamma$-ray spectrum after background subtraction.]{Logarithm scale of the $^{54}$Mn $\gamma$-ray spectrum after background subtraction. The $^{54}$Mn photopeak is at 834.848 keV. The second coincident photopeak shows at $\sim$1665 keV, and the third coincident photopeak shows at $\sim$2497 keV.}
\label{mn_FullSpectrum}
\end{figure}

With the background correction completed, the pile-up correction can be made. Figure~\ref{mn_FullSpectrum} shows the full $^{54}$Mn $\gamma$-ray spectrum after background subtraction. The $^{54}$Mn spectrum consists of a "Compton" region in which the full energy of the $\gamma$-ray was not completely absorbed. This region is labelled Region 1 in Figure~\ref{mn_FullSpectrum}. The full-energy peak, called the photopeak, is produced by the complete absorption of the $\gamma$-energy, as shown in Region 2. The ratio of events in Region 1 to Region 2 is nearly 3. One of the major systematic error in $^{54}$Mn spectrum is the electronic pile-up. The pile-up, shown in Region 3, and 4, is caused by two independent nuclear decay photons which interact with the detector within a time period shorter than the resolving time of the detector. Because the primary photopeak can be treated as a $\delta$-function, the pile-up Region 3, and 4 appears as an integration of the "Compton" Region 1, with the photopeak, Region 2, yielding a mirror-like image of the lower energy single, $\gamma$-ray region, but at higher energy. 
Region 3, and 4 is called the first pile-up of the $^{54}$Mn spectrum that has the ratio of $10^{-2}$ to the 1st photopeak. Region 5, and 6 are called the second pile-up of the $^{54}$Mn spectrum in which 3 $\gamma$-rays interact with the detector in a time shorter than the resolving time. This region has a ratio to the primary photopeak of nearly $10^{-5}$. The expected measuring sensitivity in the experiment is to be 1 part in $10^{5}$, so the pile-up corrections play a key role to improve the measuring sensitivity.\\
\indent
To obtain a highly accurate single-events $^{54}$Mn spectrum, the piled-up events are removed through an iterative de-convolution algorithm. This procedure is performed for the energy-calibrated spectra, and background subtracted spectra. Each de-convolution cycle starts with only the energy region containing the photopeak, and below. It proceeds by calculating the destination bin of any two events residing in the single-events region. As a metric, the reduced S-value \cite{York200472} of the residuals between the input spectrum to the resulting de-convolution in the first-order pileup region is computed. The de-convolution spectrum is generated in the following way,
\begin{equation}
P_{k}(E_{i}+E_{j})=\sum_{k=i+j}P_{i}(E_{i}) * P_{j}(E_{j})
\end{equation}
where it is assumed that the energy value refers to the bin center. Unfortunately, the energy associated with the convoluted bin $k$ does not match the energy bins associated with the original energy spectrum because of the nonlinear calibration. The de-convolution spectrum must be appropriately energy scaled before it can be subtracted from the original energy spectrum.
\begin{figure}[t!]
\begin{center}
\includegraphics[scale=.35]{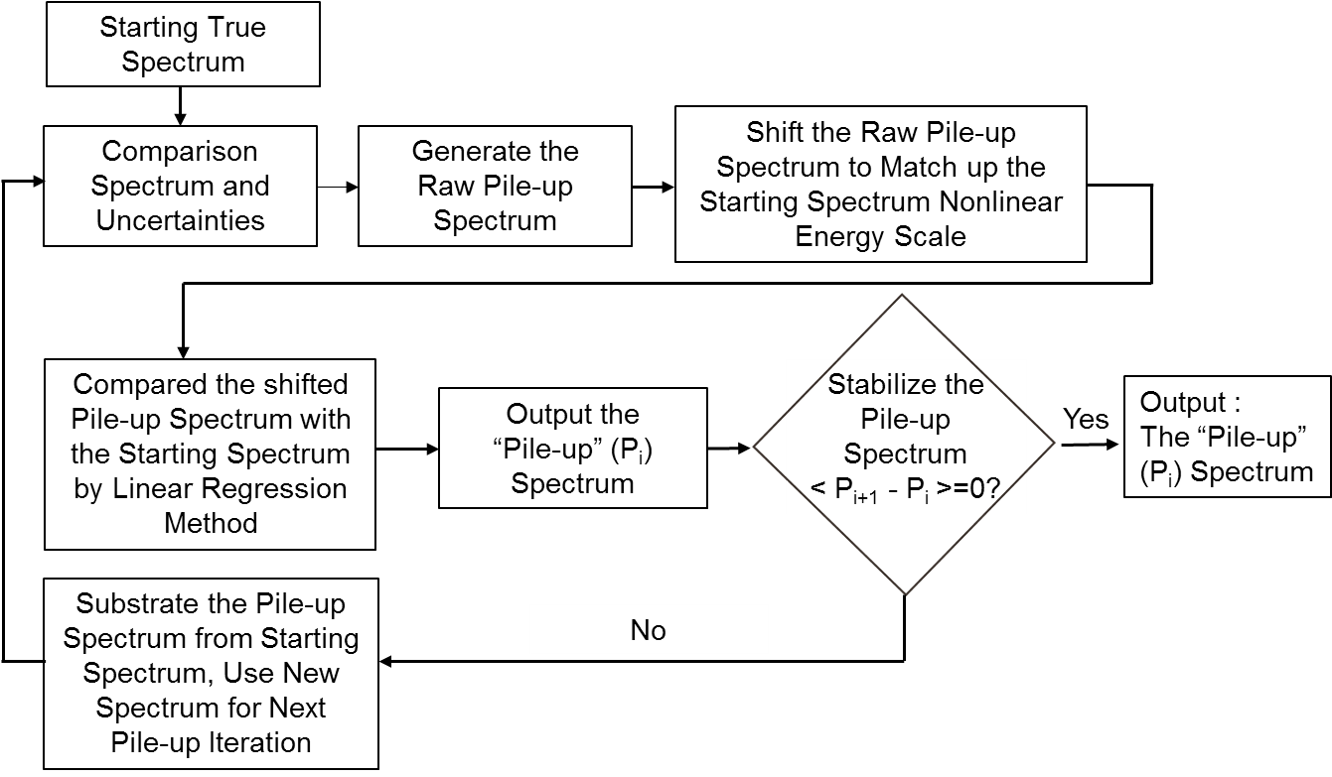} 
\end{center}
\caption{De-convolution algorithm used to determine pileup events to be removed from the $^{54}$Mn spectrum.}
\label{pile_up_algorithm}
\end{figure}
Thus, the energy scale of the pile-up spectrum must be corrected to match the original spectrum energy scale. Because the energy scale of the $^{54}$Mn spectra is non-linear, it is convenient to match up energy scales directly for this subtraction process. 
Again, only the pile-up spectrum is manipulated. This procedure is similar to the previous background correction process.\\
\indent
Once the pile-up spectrum is generated, and appropriately energy
matched to the $^{54}$Mn spectrum, the pile-up spectrum is then compared with the 1st pile-up region input the spectrum for amplitude rescaling. The comparison region for the source, and the pileup spectrum is the energy region between the end of the primary photopeak (1st), and the end of the pileup photopeak (2nd) which is within the boundaries of Region 2 to Region 4 in Figure~\ref{mn_FullSpectrum}. A linear regression method with uncertainties in 2-dimension is utilized to obtain the scaling factor\cite{York200472}. After the pile-up spectrum amplitude scaling, the pileup spectrum is subtracted from the original starting energy spectrum that served as the first guess to generate the pileup spectrum.
\begin{figure}[t!]
\begin{center}
\includegraphics[scale=.36]{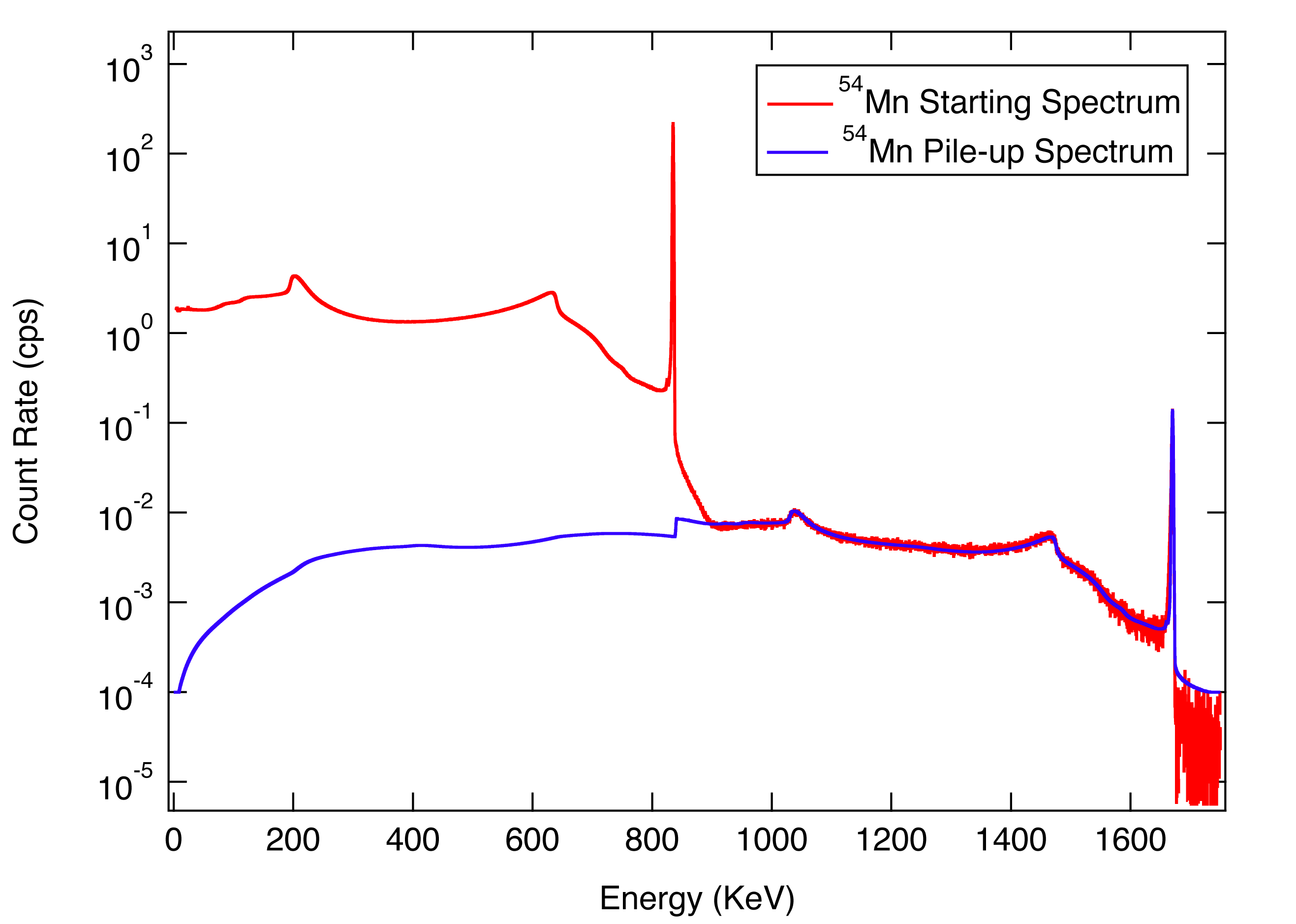} 
\end{center}
\caption{Calculated pileup spectrum (blue) with the starting spectrum (red) demonstrating the fit achieved by the de-convolution algorithm.}
\label{PileUp}
\end{figure}
This process is then iterated until the starting spectrum,  and the generated pileup spectrum stabilize. A reduced S-value test is used to stabilize the pile-up spectrum\cite{York200472}. Each iteration requires reconstruction of the de-convolution or pileup spectrum to correctly compare it to, and subtract it from its generating spectrum. The de-convolution algorithm flowchart is shown in Figure~\ref{pile_up_algorithm}. Figure~\ref{PileUp} shows the starting spectrum, and the results pile-up spectrum. As shown the method compares well in the comparison region, however, the pile-up photopeak's width is underestimated because the energy resolution is not a linear function of energy. Nonetheless, this effect does not alter the estimate of the pile-up within the region of the single event spectrum due to its smoothness in this region.\\
\indent
The $\chi^{2}_{Dof}$ of the difference between the convolution, and the data in the match region is $1.26$ indicating an acceptable convergence. Using the first day's data in which the pileup is largest, the rate difference between the convolution spectrum, and the $^{54}$Mn spectrum in the match region is $\Delta R = 8 \times 10^{-6}$ cps. Again, showing an excellent match for the convolution, as the total rate in this region is $15.176 \pm 0.014$ cps.  Because the total rate difference is so much smaller than the regions total rate error, this indicates that the error generated by the convolution method is dominated by the statistics in the $^{54}$Mn spectrum.\\
\indent
The accuracy of the background match region rate is known to $\sim 10^{-3}$. Assuming a similar background pileup rate in the region including, and below the photopeak in the $^{54}$Mn spectrum means the pileup correction is at the level of $2 \times 10^{-3}$, given the rate in the same region of the $^{54}$Mn spectrum is $\sim  8 \times 10^{3}$ cps. With these considerations, it is shown that the error generated by the convolution process is order $2 \times 10^{-6}$ which can be neglected in this analysis.\\
\indent
After energy scaling, and convergence, the pileup spectrum is subtracted from the starting spectrum generating the true $^{54}$Mn pile-up corrected spectrum.
\subsection{Region of Interest (ROI) of $^{54}$Mn Spectrum}
\begin{figure}[t!]
\begin{center}
\includegraphics[scale=.35]{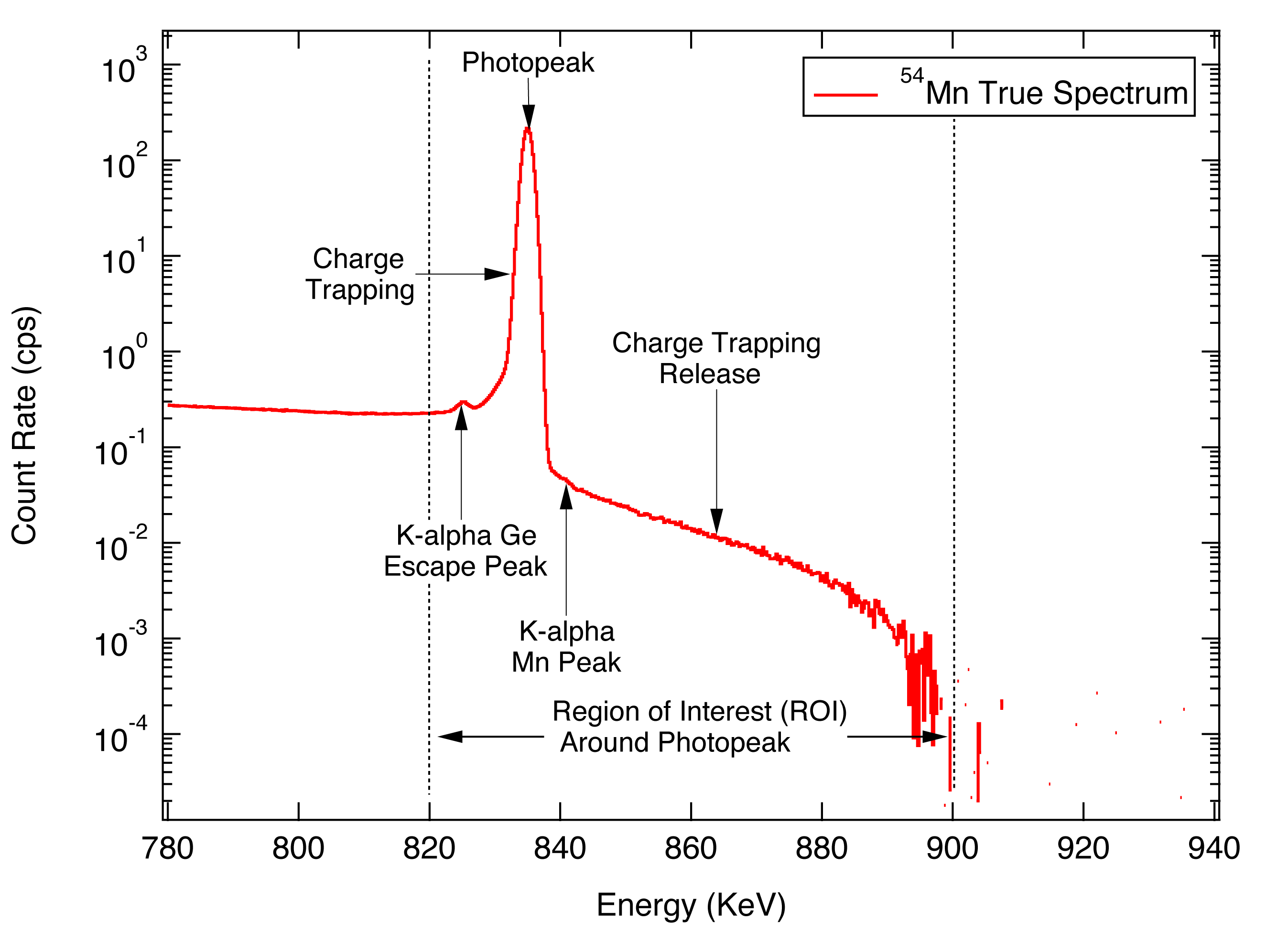} 
\end{center}
\caption{The Region of Interest, from 820 to 900 keV, is selected from the corrected daily $^{54}$Mn spectrum.}
\label{ROI_Spectrum}
\end{figure}

The daily decay rate from the $^{54}$Mn corrected spectrum can be determined using different techniques. One technique is to fit the photopeak shape. The drawback of this method is the photopeak is too complicated to model to the level of accuracy required. For example, charge trapping, and  trapped charge release occur in the HPGe detector during the $\gamma$-ray interaction process. These effects alter the HPGe line shape, as shown in Figure~\ref{ROI_Spectrum}, and may cause reduced or excess energy to be measured in the HPGe detector that is deposited by the $\gamma$-ray. As is shown in Figure~\ref{ROI_Spectrum}, the line shapes are far from having a Gaussian distribution, and no simple model can properly describe all the shape effects.\\
\indent
In this experiment charge trapping release alters
the measured $\gamma$-ray energy by up to 60 keV away from the photopeak centroid. This charge trapping is due to defects in the crystal caused by its previous exposure to neutrons. These charge release events are extremely rare.  The quality of this detector can be evaluated using the typical measure of merit, the Gaussian line width $\sigma_{E}$. The fit line width at the $^{54}$Mn photopeak is measured to be very acceptable, $\sim 1$keV.\\
\indent
However, because of the very high precision required in this experiment, $10^{-5}$, the additional effects of charge trapping release were only observable after background, and pileup corrections were made to the spectra. This effect has not been reported in the literature.  This effect has been studied\cite{CARRI}, and its observation is due to the strength of the photopeak. This feature is understood as follows; Using the first day’s spectrum in which the effect is largest, the total rate of charge release events is $3.5$ cps. This is a factional rate of $2\times10^{-3}$ when compared to the photopeak. The side band region of equal energy width below the ROI has a total rate of 0.1 cps yielding a charge release rate of $2\times 10^{-4}$ cps, which is not measurable in this experiment. Thus, the observation of the effect is due totally to the strength of the photopeak. For this reason, the ROI has been selected to fully include it as the upper end of the ROI energy band. Note that after the background, and pileup corrections, there are effectively zero events above the ROI energy band. In addition, the low energy band has been selected to include charge trapping as well as the k-alpha escape peak in the Ge crystal. These escape events occur when a surface Ge atom radiates a k-alpha photon away from the crystal. Thus, its energy is lost. Likewise, the k-alpha pileup from the $^{54}$Mn source, and the photopeak have not been removed, because these are photopeak events, while energy shifted, they are not lost out of the ROI region as is shown in Figure~\ref{ROI_Spectrum}.\\
\indent
By selection of an ROI band instead of what is an ill-defined and elusive definition of photopeak by fitting, the effects of charge trapping, and charge trapping release, and other effects do not impact the sensitivity of the measurement. Thus, the ROI in this measurement extends from 820 keV to 900 keV.  In addition, the ROI is a fixed energy region, instead of a fixed channel range. Selecting a fixed energy region eliminates the effects of calibration drift, and is accounted for by the time-dependent energy spectrum calibration coefficients.\\
\indent
Having selected the ROI, Table~\ref{limit_table} provides a summary of the error for each 24 hour data point, including the electronic dead time correction, the background spectrum subtraction, the pile-up spectrum correction, statistical error from counting, and systematic error from the instruments. The total error per day is $0.022$ cps. Averaging the daily measured decay rate in the ROI over the full length of the experiment yields on average count rate.  Using this average, the per day sensitivity in $\delta \lambda /\lambda \sim 1.5\times 10^{-5}$.
\begin{table}
\caption{ The average individual error contributions in the Region of Interest (ROI) using the $^{54}$Mn daily spectra.}
\begin{center}
\begin{tabular}{ccc}
\hline \hline
Averaged Correction  & Count Rate  & Error Rate \\ 
per Day in ROI & in ROI (cps) &  in ROI (cps) \\
\hline 
Statistical  & 1431.43 & 1.0E-02 \\
Dead time (0.029 sec/day) & 1431.43 & 9.7E-03 \\
ROI uncertainty (2.2E-2 keV) & 0.18 & 1.8E-02 \\
Background reactor-off & 0.056 & 1.4E-05 \\
Background reactor-on & 0.086 & 2.4E-05 \\ 
Pile-up  & 1.70 & 3.4E-05 \\ \hline
Total error rate (cps) &  & 2.2E-02  \\ 
Sensitivity $\delta$R/R && 1.5E-05\\
\hline \hline
\end{tabular}
\label{limit_table}
\end{center}
\end{table}

\section{Analysis of Corrected $^{54}$Mn spectra }
\begin{figure}[b!]
\begin{center}
\includegraphics[scale=.3]{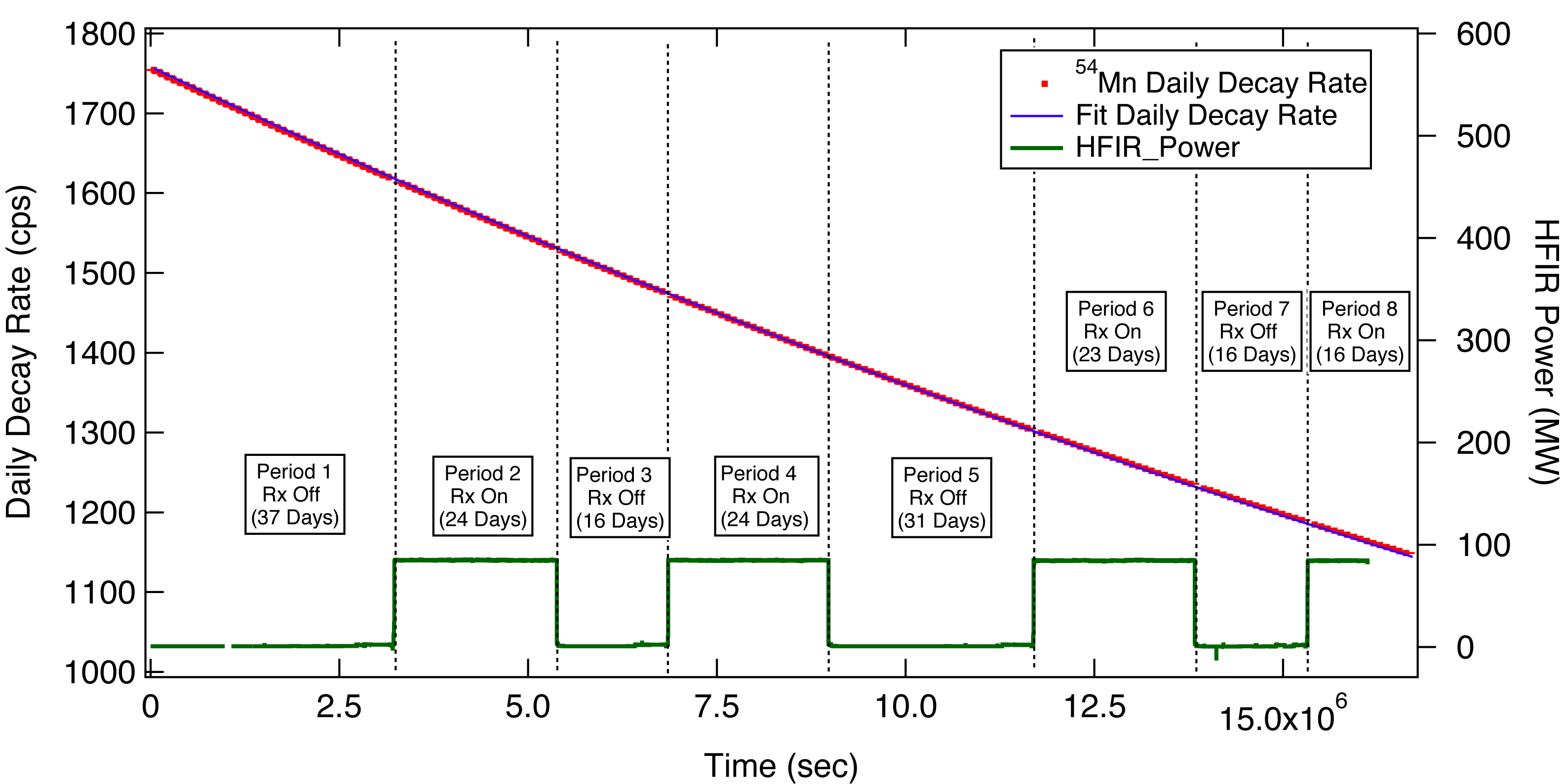} 
\end{center}
\caption[Daily $^{54}$Mn ROI (between 820 to 900 keV) decay rate.]{Daily $^{54}$Mn ROI (between 820 to 900 keV) decay rate. Error bars are too small to be shown. The HFIR reactor power is also shown. The Periods 1, 3, 5, and 7 are the reactor-off periods. The Periods 2, 4, 6 ,and 8 are reactor-on periods.}
\label{54Mn_daily_rate}
\end{figure}

The $^{54}$Mn ROI daily decay rate as a function of time is shown in Figure~\ref{54Mn_daily_rate}. The rate was calculated for each spectrum starting August 30, 2015, at 23:58:22, and ending March 09, 2016 at 23:12:51. The time period includes four reactor-off, and four reactor-on periods. 
Those daily runs which cover the transition time of the reactor-on or reactor-off period have been removed to reduce possible error. The data collection consists of a continuous period including 87 reactor-on days, and 105 reactor-off days over 192 days of data collection. The ROI count rate was initially $1754.31 \pm 0.15$ cps but decayed to approximately $1146.37\pm 0.12$ cps at the end of the experiment. The data point for each day is plotted at the average time weighed by an exponential function calculated using the standard $^{54}$Mn mean lifetime $\tau$ (450.41$\pm$0.29 days)\cite{NNDC_54MN}.\\
\indent
The analysis of the decay rate requires correction for the effect of environmental influences, as measured and discussed concerning the calibration b-term and c-term.  To correct for the b-term the  decay function includes a single periodic function with a phase to take into account the known yearly environmental oscillations observed in the data,
\begin{equation}
R(t)=  a e^{-t/\tau} + A sin(\omega t + \phi).
%b (cos(\omega t)) + c (sin (\omega t))
\label{decay_Eq}
\end{equation}
Again, $\tau$ is fixed at the mean lifetime of $^{54}$Mn\cite{NNDC_54MN}, 
$\omega$ is the periodicity, fixed at one year, and $\phi$ is the phase relative to the start of the experiment. Including the yearly environmental effects in the fit reduces the $\chi^{2}$ per degree of freedom from $\sim$100 to $1.54$, using 4-degrees of freedom. As shown in Table~\ref{Amp_phas} the the oscillation has a good match to the $b_{r}$ parameter oscillation, and has a good out of phase match to the humidity. This is as expected if the oscillation is driven by the yearly variations in the humidity acting on the X-cooler. The amplitude of the oscillation is $1.55\pm0.01$ cps. When compared to the average rate in the ROI, $1431.43\pm0.18$ cps the fractional effect is at the level of $1.1\times10^{-3}$.\\

\begin{table}
\caption{Comparison of in-phase, and out-of-phase ROI oscillation.}
\begin{center}   
\begin{tabular}{c c|c c}
\hline \hline
ROI Oscillation & Phase ($\phi$) & In phase & Phase ($\phi$)  \\
\hline
In phase  & $-111\pm0.5$ day& $b_{r}$ terms  &$-122 \pm 2$ day  \\
Out of phase & $71\pm0.5$ day  & Humidity & $79\pm1$ day \\
\hline\hline        
\end{tabular}       
\end{center}
\label{Amp_phas}
\end{table}
 
\begin{figure}
\begin{center}
\includegraphics[scale=.35]{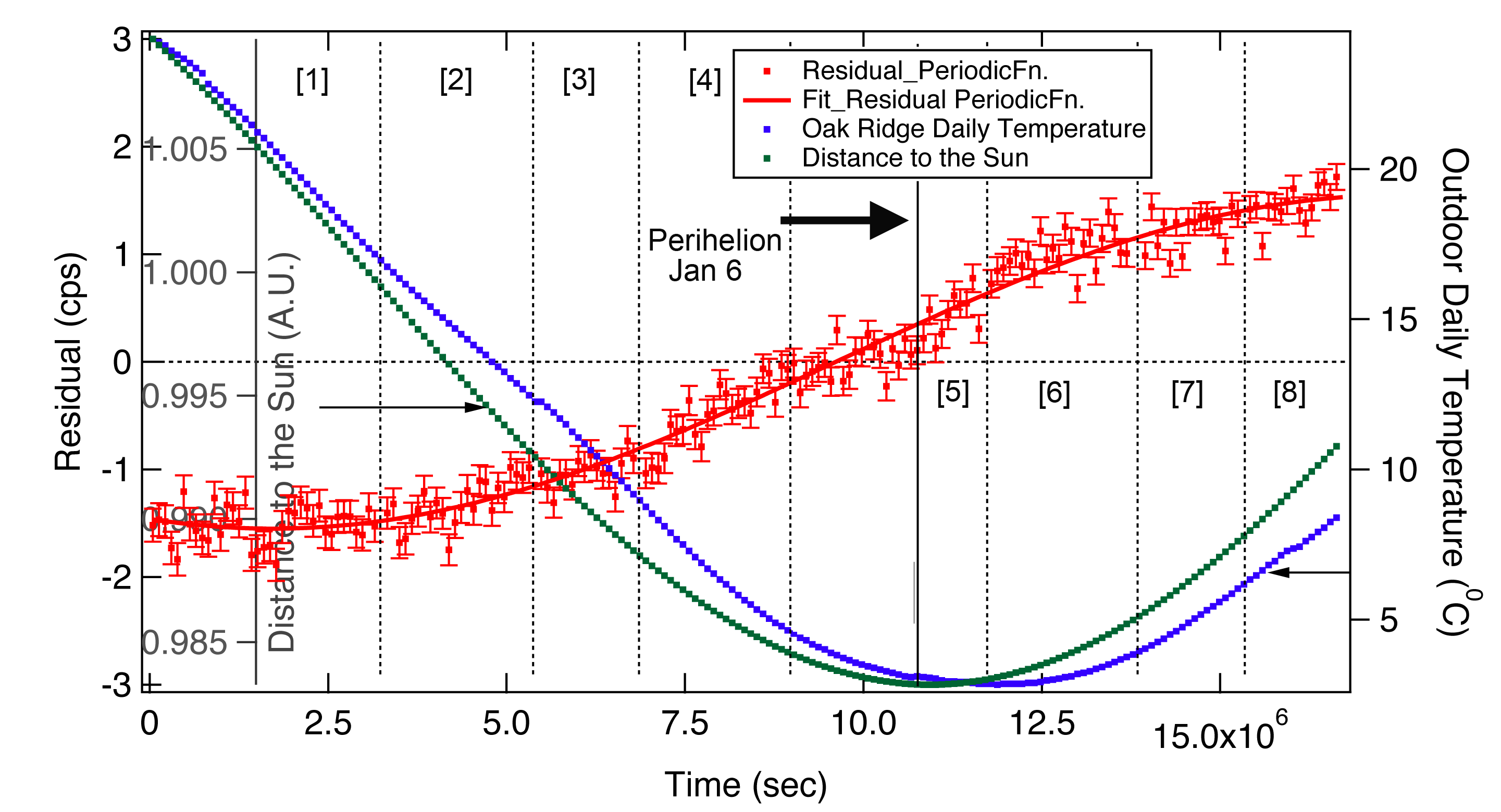} 
\end{center}
\caption[The oscillation term found by subtraction of the exponential term from the data (Eq.~\ref{decay_Eq}), and its fit shown in Red data points]{The oscillation term found by subtraction of only the exponential term from the data (Eq.~\ref{decay_Eq}), and its fit shown in Red data points. For comparison, the daily outdoor temperature, and the Earth's distance from the Sun are not in phase with the environmental oscillations inside the HFIR complex.}
\label{54Mn_res_ExpFourier_2}
\end{figure}
\indent
The fitting function, Eq. \ref{decay_Eq}, successfully removes the oscillation behavior. The minimum value occurs on Sep/20/2015 which is nearly 21 days from the starting date Aug/30/2015.  This date is not associated with the Earth's perihelion or the aphelion which occurred Jan/02/2016 17:49 (EST), and Jul/06/2015 14:40 (EST)\cite{usno}, as shown in Figure~\ref{54Mn_res_ExpFourier_2}. If the periodicity is allowed to vary, the $\chi^{2}/DoF$ is unchanged, and yields an oscillation of 363.6 days, in agreement within error, with the 1-year fixed value. Therefore, the oscillation of the decay rate does not correlate to the solar neutrino flux variation due to the Earth's motion.\\
\indent
As an aside, The subtraction of this low-frequency term in no way affects the sensitivity of the search for decay rate parameter variations in this experiment, but instead demonstrates the ability to reject environmental effects, as the HFIR characteristic on-time period is 30 days a much higher frequency than the 1-year environmental frequency being filtered-out in this search.

\section{Side Band Nonlinear Energy Calibration Corrections due to Environmental Effects}
\begin{figure}[t!]
\begin{center}
\includegraphics[scale=0.37]{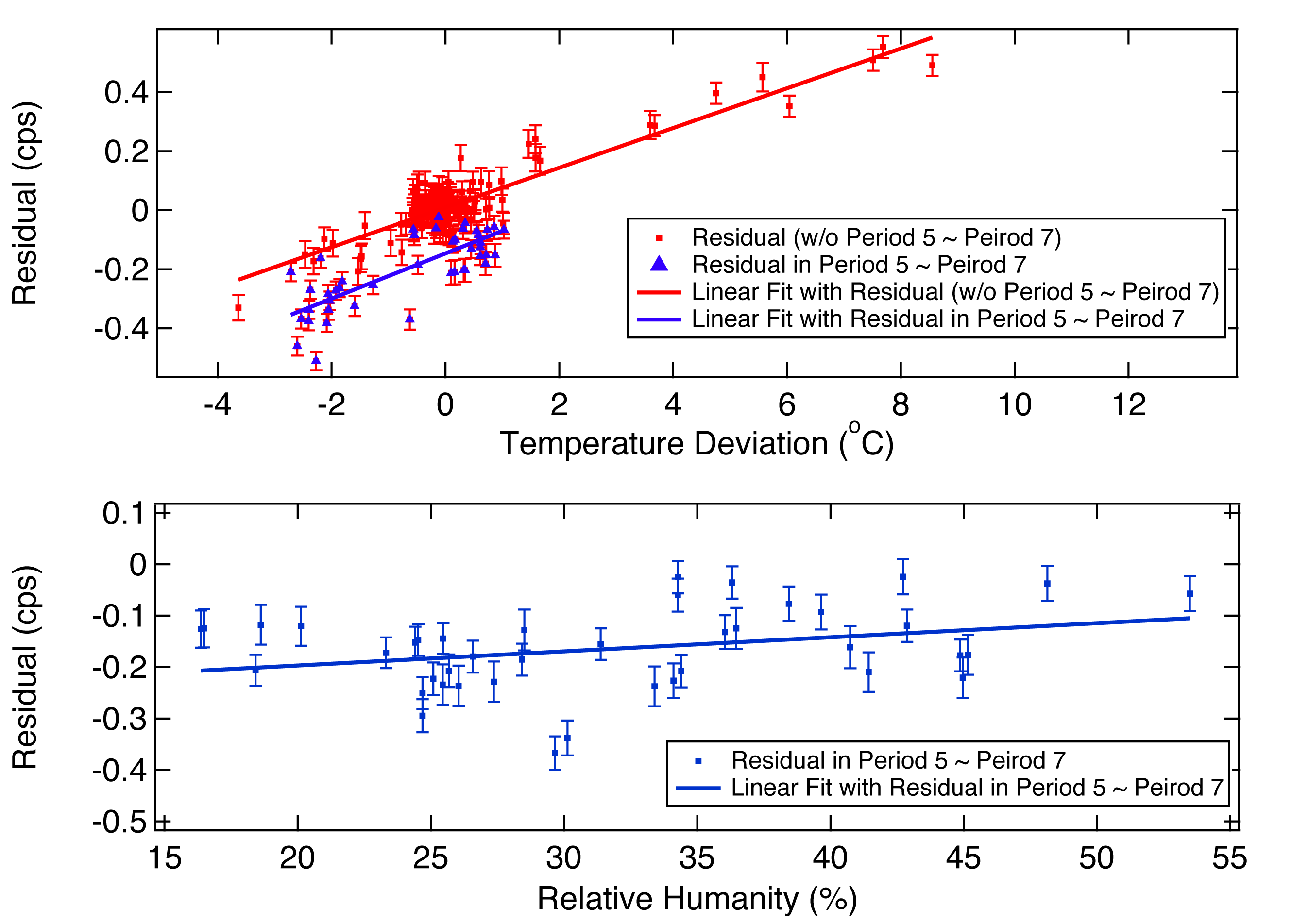} 
\end{center}
\caption{(Upper)Daily residuals of the side band energy region as a function daily average Temperature.(Lower) Daily residual of the side band energy region as a function of daily relative humidity in Periods 5, 6, and 7 during periods having near zero temperature variations.}
\label{Temp_RH}
\end{figure}

As with the linear calibration term, because the DEPEC-50  spectrometer electronics and the X-cooler are located outside the controlled environmental housing, environmental factors such as temperature, pressure, and humanity cause variations in the nonlinear energy scale calibration parameter c$_r$. The use of nonlinear calibrations has previously been presented for both the source, and background spectra. The nonlinear energy scale correction is given in Eq. \ref{eq:cal_param} and repeated here for clarity,
\begin{equation}
E = a_{r} + b_{r}x + c_{r}x{^2}
\label{eq:cal_param_2}
\end{equation}
where again, $a_{r}$, $b_{r}$, and $c_{r}$ are the calibration coefficient of $r$th daily spectrum, and $x$ is the channel number of the spectrum. The ROI data is analyzed using the counts per second (cps) in a fixed energy band $\Delta E$ which includes the photopeak. For illustration, using a band containing n fixed bins, this band is given by
\begin{equation}
\begin{split}
\Delta E & = b_{r}({x_{i+n}} - {x_{i}}) + c_{r}({x_{i+n}^{2}} - {x_{i}^{2}})\\
         & = \left[ {b_{r} + c_{r}({x_{i + n}}{ + x_i})} \right] \times ({x_{i + n}} - {x_i})\\
\end{split}
\end{equation}
The error in the bands energy width generates an error in the counts associated with that band. The error in the width is
\begin{equation}
    \delta (\Delta E) = \left[ \delta b_{r} + \delta c_{r}(x_{i+n} + x_{i} ) \right] \times (x_{i+n}-x_{i})
\label{deltaE1}
\end{equation}
$b_{r}$ is well measured to a fractional accuracy of $7\times10^{-5}$, as shown in Figure~\ref{Nonlinear_5lines_abc}. $c_{r}(\sim10^{-8})$ is measured to a fractional accuracy of order $10^{-1}$. Nonetheless it is sensitive to environmental effects.  
\begin{figure}[t!]
\begin{center}
\includegraphics[scale=.51]{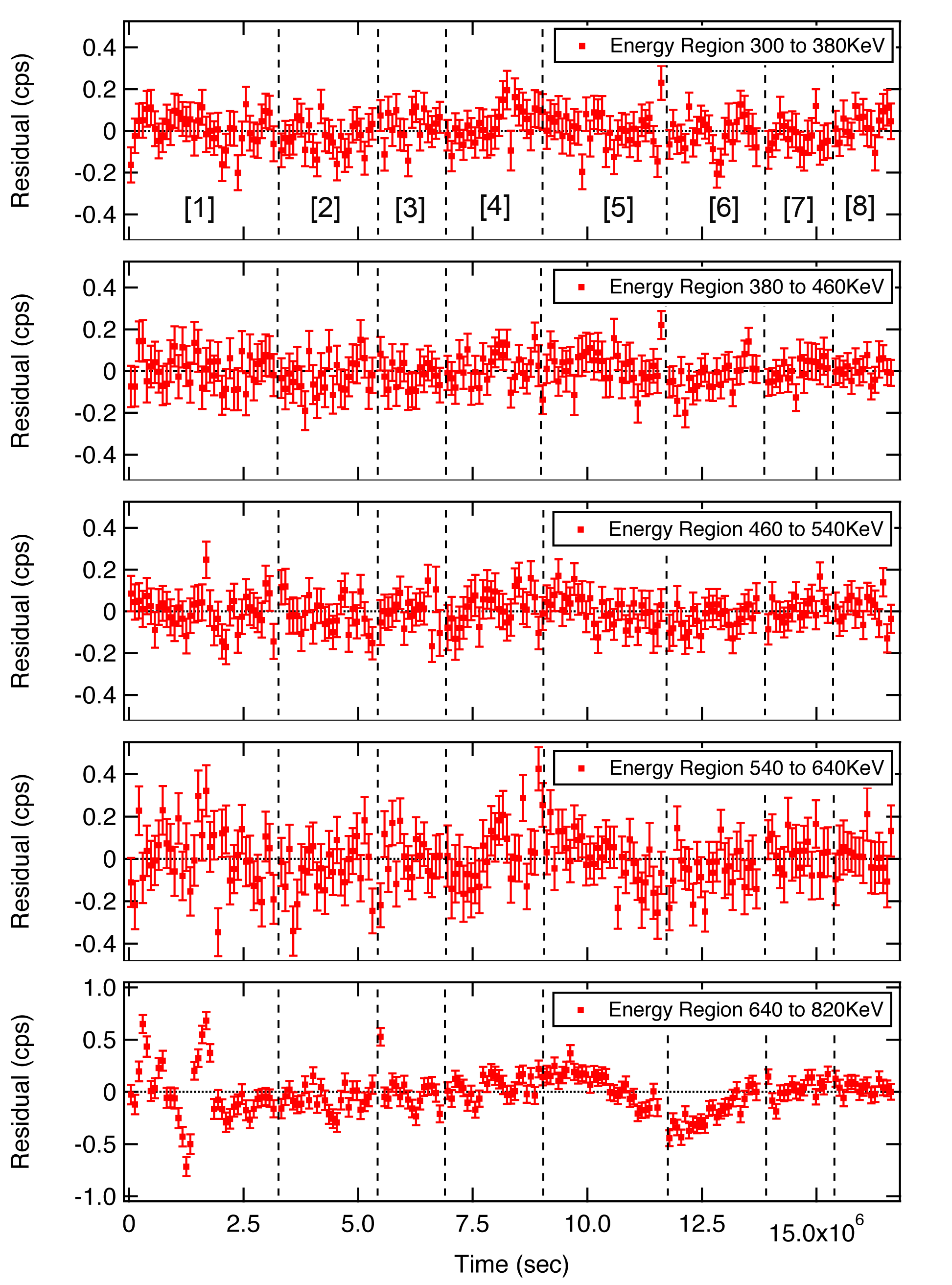} 
\end{center}
\caption[Residuals of equal width energy regions from the corrected $^{54}$Mn spectrum.]{Residuals of equal width energy regions from the corrected $^{54}$Mn spectrum. The residuals show the environmental effects strongly correlate with the increasing strength of the nonlinear term $c_{r}x^{2}$.   Band in keV, (Form the Top), (300,380), (380,460), (460,540), (540,640), (640, 820).}
\label{Energy300_820KeV}
\end{figure}
At low bin number the non-linear term has no effect on the energy band width. However, as the bin number increases, and in the ROI band where $x_{i} \sim 4000$, the error in the band width is dominated by the non-linear term 
\begin{equation}
   (\delta {c_r} \cdot 2{x_i})^{2} \ge {(\delta {b_r})^2}
\end{equation}
so that the error on the width is given approximately by
\begin{equation}
    \delta (\Delta E)\sim \delta {c_r}(x_{i+n}^{2}-x_{i}^{2})
\end{equation}
The knowledge of $c_{r}$-term motion as a function of the environmental parameters can be found by study of the side band residuals.\\
\indent
These expectations concerning $\delta c_{r}$ are verified in the spectral data by study of the energy regions below the ROI.  Regions of equal energy width were selected staring from 300 to 380 keV, 380 to 460 keV, 460 to 540 keV, 540 to 640 keV as well as 640 to 820 keV all below the ROI. The daily decay rate for each of these energy regions is fit to the same function as the ROI, Equation~\ref{decay_Eq}. Again, only $a$, and $A$ are variables in the fit. $\omega$, $\tau$, and $\phi$ are the same coefficients used for the ROI.  Figure~\ref{Energy300_820KeV} shows the residual for each energy regions, and 
\begin{table}[t!]
\caption{The $\chi^{2}$ per degree of freedom after fitting equal width energy regions in Figure~\ref{Energy300_820KeV}.}
\begin{center}   
%\setstretch{1.25}
\begin{tabular}[b]{cc}
        \hline\hline Energy & $\chi^{2}$ per    \\ 
                Region (keV) &  degree of freedom\\
        \hline 
			   300 to 380  & $0.98$    \\        
        	   380 to 460  & $0.86$    \\     
               460 to 540  & $0.87$    \\
               540 to 640  & $1.24$    \\
               640 to 820  & $12.02$   \\
        \hline\hline        
        \end{tabular}       
\end{center}
\label{Chi_Region}
\end{table}
Table~\ref{Chi_Region} gives the $\chi^{2}$ per degree of freedom for each energy region.\\
\indent
As expected the $\chi^{2}$ per degree of freedom for the lower energy bands is $\sim 1$. However, because the $c_{r}$term has a significant effect only at high bin numbers, the $\chi^{2}$ per DoF increases in the side band region just below the photopeak to $\sim$12, due to the influence of environmental effects.\\
\indent
The environmental effects on $c_{r}$ are significant causing an incorrect assignment of the energy width of the ROI.  An incorrect energy width causes motion of events from one band to another, to be lost or gained as a function of the environmental changes. The motion of events is measured by the difference in the error in the energy width of the edge bins of the energy region. These effects can be calculated from
\begin{equation}
{R_{L}}(f{({ E_{L}})}) - {R_U}(f( E_{U})) = \delta R
\label{chi_E}
\end{equation}
where $f(E_{L})$, and $f(E_{U})$ are the fractional variation in the edge bins of the side band at the lower, and upper edge. $R_{L}$, and $R_{U}$ are the rates in the side band edge channels of the spectrum. $\delta R$ is the residual rate in that energy band.  Because events enter or leave the band only through the edges, environmental effects on $c_{r}$ can be corrected for by relating $c_{r}$ to the residuals caused by the environmental factors.
Eq.~\ref{chi_E} can be rewritten as
\begin{equation}
{R_{L}}\frac{\delta c_{r} ( x^{2}_{L})}{{\Delta {E_{L}}}} - {R_U}\frac{{{\delta c_{r}} (x^{2}_{U})}}{{\Delta {E_U}}} = \delta R,
\label{Boundary2}
\end{equation}
where $\delta c_{r}$ is the correction to $c_{r}$ found using the side band residuals where the residuals are dominated by $\delta c_{r}$, due to environmental factors. This correction provides 
\begin{figure}[t!]
\begin{center}
\includegraphics[scale=.4]{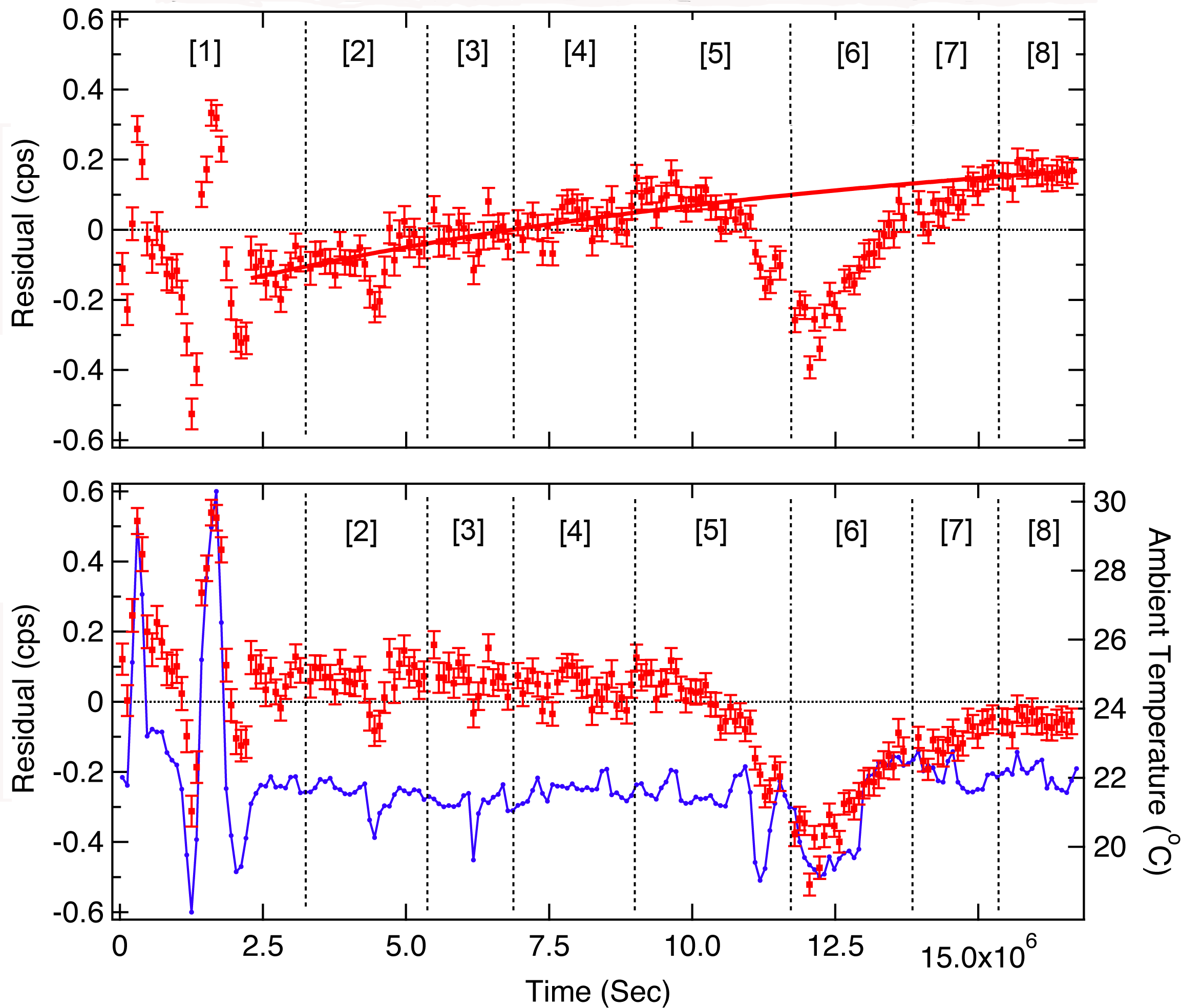} 
\end{center}
\caption{(Upper) Daily residual of the side band energy region in the $^{54}$Mn spectra. (Lower) Daily residual of the side band energy region of $^{54}$Mn spectra after the exponential function correction. A significant fluctuations appear in Period, 1, 5, and 6, due an HVAC outage within the HFIR building.}
\label{res_740820KeV}
\end{figure}
a unique way to correct small electronic effects due to environmental factors.\\
\indent
Once the corrections are found to the non-linear calibration term using the side band, this improved knowledge is used to correct the ROI band. In order to find the correction in the ROI, the same energy width band has been selected starting from 740 keV to 820 keV which is the lower side band of the ROI. The upper edge of this energy region connects to the lower edge of the ROI.\\
\indent
Before the environmental $\delta c_{r}$ corrections can be found, a false decay term they induce must be subtracted from the side band. The residuals to ROI lower side band are shown in Figure~\ref{res_740820KeV}. To correct for this, an exponential decay function, used only on this side band region, is fit to the residuals.  
\begin{equation}
\delta R(t)= C + D \times exp(-t/\tau_{1})
\label{curve}
\end{equation}
where $C$, $D$, and $\tau_{1}$ are the fitting coefficients of this function.
Once fit, the residuals in the energy region 740 to 820 keV  drop the $\chi^{2}$ per DoF from $12$ to $1.49$. $\tau_1$ is found to have low frequency,  139 days, showing this correction is not related to reactor operations, having a charactoristic frequency of $\sim$30 days.\\
\indent
After removing the lower side band false decay residual strength, Figure~\ref{Temp_RH} makes clear the strong correlation between the  remaining daily residual with temperature and humidity variations. 
The red squares in Figure~\ref{Temp_RH}(Upper) include all side band residual data points. The blue triangles only include the end of Period 5 to 7 which is related to a significant drop in temperature due to an HVAC outage in the HFIR building, effecting only the equipment outside the shielding house. It is clear there are $\pm0.6$ cps variations with temperature deviations of -4 to 8 $^{o}$C. Two linear fittings have been applied to the data set independently. The fitting coefficients from the red data point are different compared with the blue data points. The residual variations with temperature prove that environmental effects are causing the motion. However, the length of the time-dependent temperature variations also plays a role causing the difference between the red, and blue data points. Because the variations of blue data points include both reactor-off, and reactor-on periods, they are not caused by the reactor status. That is, antineutrino exposure is not the reason for this effect. \\
\indent
Likewise, the residuals motion is correlated with humidity as shown in Figure~\ref{Temp_RH} (Lower), decoupled from the temperature variations by using only those data periods having near zero temperature deviation. It should be noted that the residual shifts displayed in Figure~\ref{res_740820KeV} do not coincide with reactor-on, and reactor-off cycles. Because of the strong correlation of the side band residuals with environmental factors, these residual shifts are taken as environmental, to be used to correct for environmental factors in the ROI. That is, after the false decay subtraction, the remaining residual from the lower side band, having an energy band width equal to the ROI region, are set to zero as a measure of the environmental factors. This pre-assumes that no measurable antineutrino effects are measurable in the side band. This is reasonable as the data rate ratios between the side band $\sim 0.1$cps compared to the photopeak ROI of 1800cps is $5.6\times10^{-5}$. To further this point, if the size of the positive effect is $\sim10^{-3}$ \cite{Jenkins2009407,Jenkins200942}, as reported, then sensitivity required of the side band compared to that of the photopeak is  below the level of $\sim10^{-7}$, not measurable in this experiment.

\begin{figure}[t!]
\begin{center}
\includegraphics[scale=0.35]{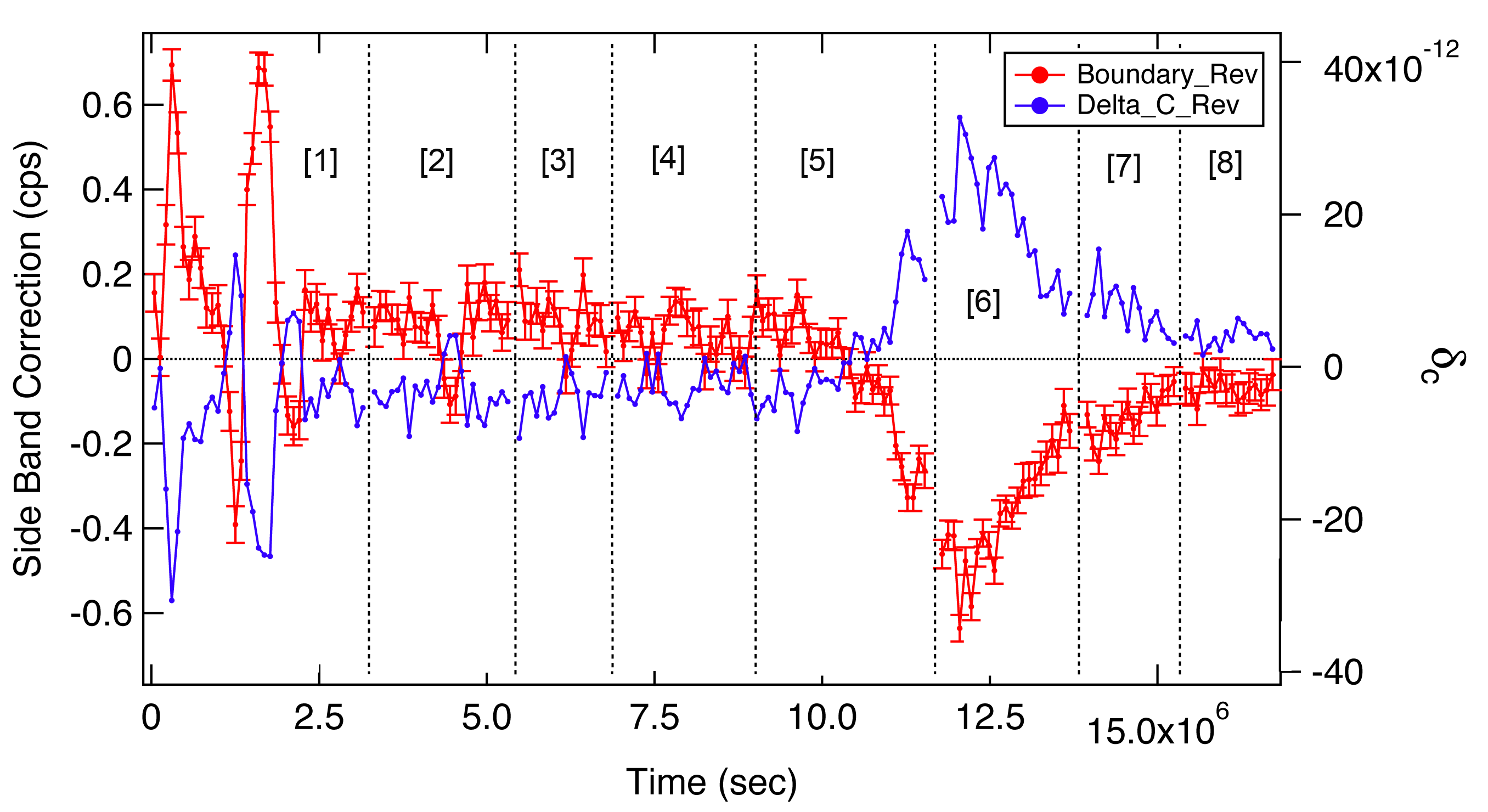} 
\end{center}
\caption{Daily $\delta c_{r}$, and corrected residual of Region 740 keV to 820 keV of $^{54}$Mn spectra, and daily average Temperature.}
\label{Delta_C}
\end{figure}
\indent
It is noted the lower edge of the ROI is the same as the upper edge of the side band edge. The upper edge of the ROI is zero after the pile-up correction. After using Eq.\ref{Boundary2} to find $\delta c$ from the side band, The corrected daily decay rate for the ROI is then,
\begin{equation}
\delta R_{ROI} = {R_U}\frac{{{\delta c_{r}} x^{2}_{U(Side\,Band)}}}{{\Delta {E_{U(Side\,Band)}}}} 
\label{Boundary3}
\end{equation}
where $R_{U(Side\,Band)} = R_{L(ROI)}$, and $x^{2}_{Side\,band} = x^{2}_{ROI}$ Note the sign changes for the correction. The rate from the upper side band edge is the lower edge of the ROI. This occurs because events lost from one band edge is a gain to the other band. \begin{equation}
R_{Corrected \, ROI}(t)=R_{ROI}(t) + {\delta R_{ROI}}
\end{equation} 
\indent
To conclude this section, it must be emphasized that the false exponential, was not used to correct the ROI or side band data sets in any way.    

\section{$^{54}$M\lowercase{n} Results After All Corrections (ROI)}
\begin{figure}[t!]
\begin{center}
\includegraphics[scale=0.25]{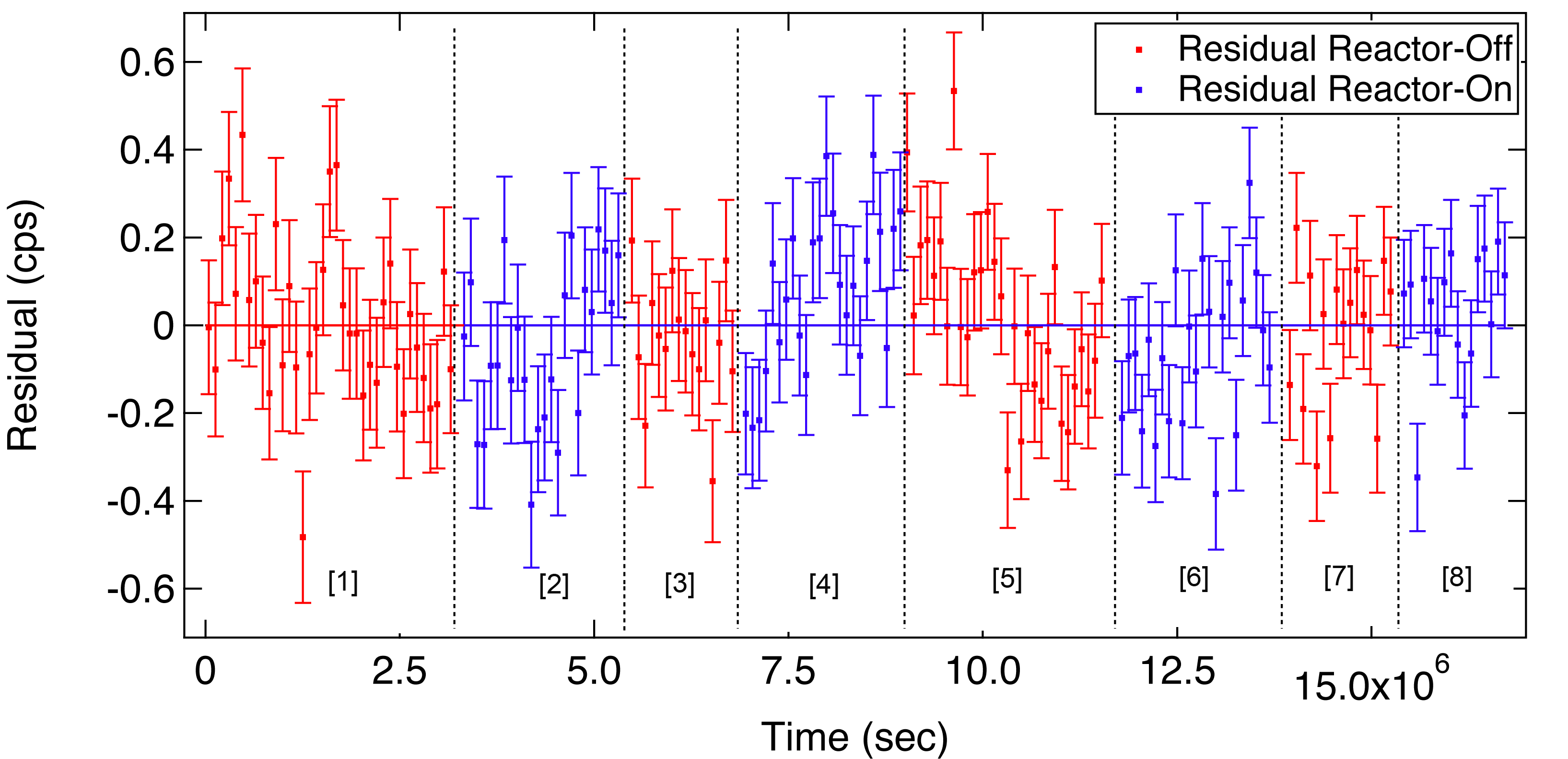}
\end{center}
\caption[Corrected ROI Residuals as a function of time.]{Corrected ROI Residuals as a function of time. The red data points indicate the residuals in reactor-off periods, and blue data points indicate the residuals in reactor-on periods.}
\label{Res_final_all}
\end{figure}
The corrected daily decay rate from the $^{54}$Mn ROI was used to search for antineutrino interactions through decay parameter variations. The corrections made to the ROI rate are (1) the background subtraction correction, (2) the pile-up correction, (3) the fixed exponential fit subtraction, (4) a fixed periodic fit subtraction (Eq.~\ref{decay_Eq}), and (5) a side band environmental correction. Figure~\ref{Res_final_all} shows the final residuals for the ROI as a function of time and reactor status. The red data points indicate the residuals in reactor-off periods, and blue data points indicate the residuals in reactor-on periods. Three different methods were used to search for an effect, and estimate sensitivity, including (1) A simple average of the residual form each period without regard to reactor status. (2) Segment analysis by checking 3 consecutive reactor cycles (3) Step search, by comparing all reactor-on residuals to all reactor-off residuals.\\
\begin{table}[t!]
\caption[Average residuals, and uncertainty of each reactor-on, and reactor-off period.]{Average residuals, and uncertainty of each reactor-on, and reactor-off period. The averaged results without regards to reactor status. The mean of all the data is less than 1-standard deviation. This is one test showing the data's consistency indicating no effect during the antineutrino exposure. $R$ is taken as the experimental average rate in ROI.}
\begin{center}   
%\setstretch{1.25}
\begin{tabular}{cccc}\hline\hline
\multicolumn{4}{c}{Average Residual Analysis}\\
\hline
\multirow{2}{*}{Period}  & Reactor & \multirow{2}{*}{Residual (cps)} & \multirow{2}{*}{Uncertainty (cps)}  \\
 & Status &  & \\ 
  \hline
 1 & Off & 9.35E-03 & 3.02E-02 \\
 2 & On & -5.01E-02 & 3.07E-02 \\
 3 & Off & -3.24E-02 & 3.57E-02 \\
 4 & On & 7.52E-02 & 2.92E-02 \\
 5 & Off & 2.24E-03 & 2.48E-02 \\
 6 & On & -5.81E-02 & 2.73E-02 \\
 7 & Off & -1.90E-02 & 3.28E-02 \\
 8 & On & 3.41E-02 & 3.22E-02 \\
\hline
\multicolumn{2}{c}{\begin{tabular}[c]{@{}c@{}}Averaged  \\   $\delta R$ (cps)\end{tabular}} & 5.18E-03 & 1.08E-02 \\
\hline
\multicolumn{2}{c}{\begin{tabular}[c]{@{}c@{}}Average Sensitivity \\     $\delta R/R$\end{tabular}} & 3.62E-06 & 7.53E-06 \\
\hline\hline
\end{tabular}
\end{center} 
\label{Effect_All}
\end{table}
\indent
A simple first test of the data is to average all residuals from each reactor-on period, and reactor-off period. This average analysis tests the consistency with the flatness of the residuals. Table~\ref{Effect_All} shows the average residual of each period yielding the size of its motion away from zero, $\delta R$(cps). Averaging all the periods without regard to reactor status yields $\delta R = (0.52\pm1.08)\times10^{-2}$cps, and $\delta R/R = (3.62\pm7.53)\times10^{-6}$ where $R$ is taken as the experimental average rate. This test shows the residuals are consistent with an origin having a single value, zero, indicating no effect during the antineutrino exposure.\\
\indent
A second method is the walking window technique, taking advantage of the alternating reactor cycle pattern for the antineutrino flux. The two like reactor status periods are averaged into a single data point, and compared to the average of the opposite reactor status they guard.  The result is $\delta R = (2.11\pm2.13)\times10^{-2}$cps, and $\delta\lambda /\lambda = (1.48\pm1.49)\times10^{-5}$. The results are within one standard deviation, and again consistent with no effect during the antineutrino exposure.\\
\indent
The strongest test of the data is to combine all the reactor-on data, and compare to all the reactor-off data in search of a step. This method yields $\delta$ R = ( 4.83 $\times$ 10$^{-4}\pm1.98\times10^{-2})$ cps and $\delta\lambda /\lambda$ = (0.033$\times$10$^{-7}$  $\pm$1.38)$\times10^{-5}$. \\
\indent
Because the walking window method creates correlations and does not use all the periods with equal strength the limits are set using the combined technique and are shown in Table \ref{Final_Upperlimit_M}.

\begin{table}[ht!]
\caption{Summary of measurements at HFIR of $^{54}$Mn decay rate variations measurement.}
\begin{center}   
%\setstretch{1.25}
\begin{tabular}[b]{cc}
\hline\hline
         Antineutrino  & \multirow{2}{*}{ $F_{\bar \nu } = 2.86 \times 10^{12}$ ($sec^{-1}cm^{-2}$) }   \\
          Flux  &     \\
        \hline
        Measured  &  \multirow{2}{*}{$\dfrac{\delta \lambda}{\lambda} = (0.034\pm 1.38)\times 10^{-5}$} \\
        Variation&  \\
        $68\%$ Upper Limit & \multirow{2}{*}{ $ \dfrac{\delta \lambda}{\lambda}  \leq 1.41\times10^{-5} $ }   \\
         Confidence Level &     \\
        \hline
       Measured Cross Section  & \multirow{2}{*}{$\sigma = (0.031\pm1.24)\times 10^{-25}$ ($cm^{2}$) }  \\
       Sensitivity &  \\
        $68\%$ Upper Limit & \multirow{2}{*}{$\sigma \leq 1.29\times10^{-25}$ ($cm^{2}$) }  \\
        Confidence Level&  \\
        \hline\hline        
        \end{tabular}       
\end{center}
\label{Final_Upperlimit_M}
\end{table}

\section{$^{137}$C\lowercase{s} Analysis}
\begin{figure}[t!]
\begin{center}
\includegraphics[scale=0.5]{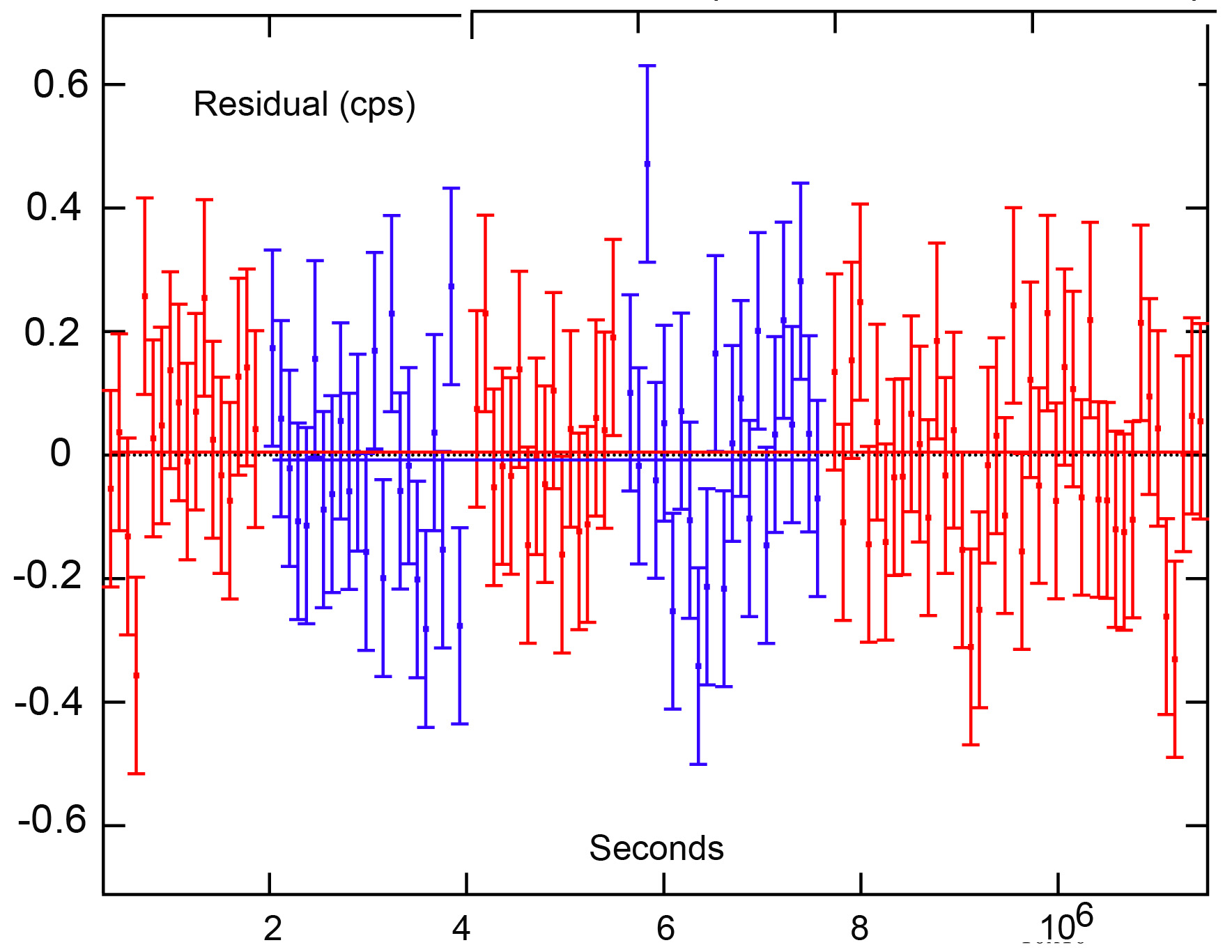} 
\end{center}
\caption{ $^{137}$Cs ROI Residuals as a function of time, after corrections (1), (2), and (3). The red data points indicate the residuals in reactor-off periods, and blue data points indicate the residuals in reactor-on periods.}
\label{Cs_final_residuals}
\end{figure}

Once data collection was complete using the $^{54}$Mn source it was exchanged for a 1 $\mu$Ci $^{137}$Cs disk source.  Data collection commenced 5-July-2016 at 13:07 EST, when the enclosure and HPGe crystal returned to thermal equilibrium.  Data collection concluded 13-November-2016.  During this interval, HFIR was at full power for 51 days during two reactor on cycles with refueling outages spanning 77 days in aggregate. Again, data was collected in energy bins $\sim$0.225 keV in width, over the energy region from $\sim$4 keV up to $\sim$3.7 MeV. During a 24-hour data run approximately 6.2 x 10$^8$ counts were collected with an associated statistical uncertainty of 4 parts in 10$^5$.\\
\indent
The $^{137}$Cs photopeak is at 661.67 keV and has a 30.03$\pm$0.09 year half-life. The ROI was selected to extend from 640 keV to 680 keV.  During the course of the 128 day experiment the ROI count rate of the source ranged from 1.8996kHz to 1.8846 kHz. Both this longer half-life and lower energy photopeak allowed the analysis of possible antineutrino effects on its decay to be considerably simpler in comparison to $^{54}$Mn.\\
\indent
The corrections made to the $^{137}$Cs ROI rate were the same extensively discussed in the analysis of the $^{54}$Mn spectra. The corrections to the $^{137}$Cs spectra included, (1) the background subtraction correction.  The background in the ROI was measured to be 0.064$\pm$0.001Hz when the reactor was on,  and 0.044$\pm$0.001Hz when the reactor was off. (2) The pile-up correction.  The estimated remaining count rate, $\sim$0.035 Hz in each fully corrected spectrum, was estimated by  extrapolation of the remaining counts in the pile-up region into the corrected region.  This results gives a full spectrum pileup error of $\sim5\times10^{-6}$. And, (3) the fixed exponential fit subtraction. The residuals after corrections (1),(2), and (3) are displayed in Figure \ref{Cs_final_residuals} and when comparing the reactor-on to reactor-off periods, yield a variation of $\delta\lambda/\lambda = (0.67\pm1.56)\times10^{-5}$. The 4$^{th}$ correction used for the $^{54}$Mn spectra, the fixed periodic fit subtraction, had little effect on the residuals, yielding the result, $\delta\lambda/\lambda = (1.68\pm1.56)\times10^{-5}$.  However, unlike the  $^{54}$Mn fit using Equation \ref{decay_Eq}, when $\omega$ was unfixed the fit returned $\omega$=0. That is no oscillation. This is reasonable because of the shorter running period of the $^{137}$Cs running period in comparison to $^{54}$Mn and that the $^{137}$Cs running period was contained in a single season in which environmental parameters varied little.    In addition, the the 5$^{th}$ correction used for the $^{54}$Mn spectra, the side band environmental correction  was unnecessary. For  $^{137}$Cs, due to the  lower energy ROI, the motion caused by the nonlinear term, as shown in Table \ref{Energy300_820KeV} for $^{54}$Mn,  is contained within the ROI band so has no effect on the residuals.\\
\indent
The $68\%$ confidence upper limits for antineutrino interaction on $^{137}$Cs, using corrections (1), (2), and (3) are displayed in Table \ref{Final_Upperlimit}.

\begin{figure}[t!]
\begin{center}
\includegraphics[scale=0.55]{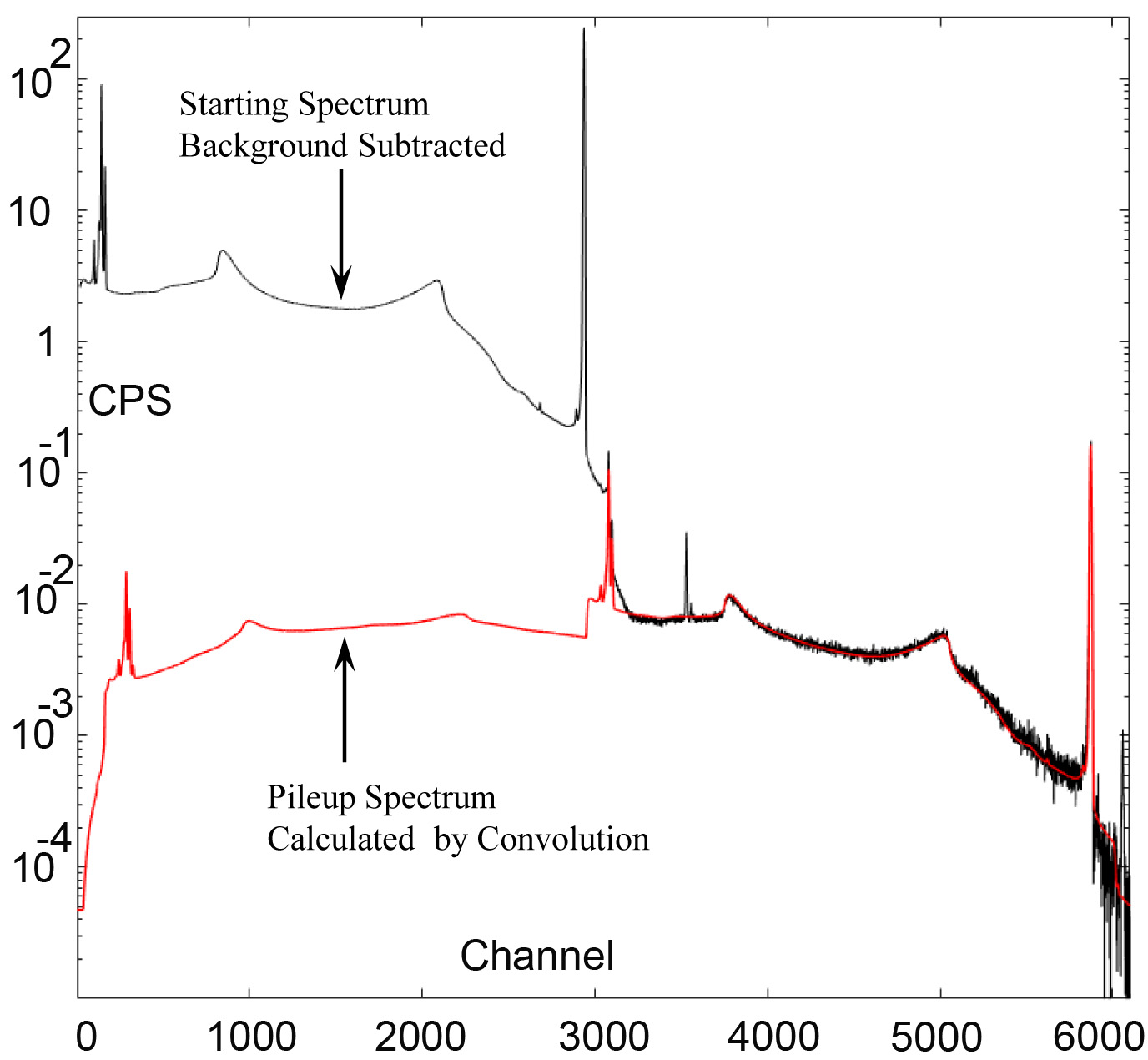} 
\end{center}
\caption{$^{137}$Cs pileup correction for the first daily run spectrum. Black: The uncorrected spectrum.  Red: The pileup background to be subtracted.}
\label{Cs_pileup}
\end{figure}

\begin{table}[htb!]
\caption{$68\%$ confidence upper limit on antineutrino interaction on $^{137}$Cs}.
\begin{center}   
\begin{tabular}{cc}
\hline\hline
        Measured  & \multirow{2}{*}{ $ \dfrac{\delta \lambda}{\lambda} = (0.67\pm 1.56)\times10^{-5} $ }   \\
         Variation &     \\
           68\% Upper Limit  & \multirow{2}{*}{ $ \dfrac{\delta \lambda}{\lambda} \leq 2.23\times10^{-5} $ }   \\
         Confidence Level &     \\
        \hline
        Cross Section& \multirow{2}{*}{$\sigma = (1.71\pm 3.98)\times10^{-27}$ cm$^2$ }  \\
        Sensitivity&  \\
         68\% Upper Limit  & \multirow{2}{*} {$\sigma \leq 5.69\times10^{-27}$ cm$^2$}   \\
         Confidence Level &     \\
        \hline\hline        
        \end{tabular}       
\end{center}
\label{Final_Upperlimit}
\end{table}

\section{Conclusion}
The experiment has placed limits on decay rate parameter variation with sensitivity at 1 part of $10^{5}$ by measuring the $\gamma$ spectra from $^{54}$Mn electron capture decay and $^{127}$Cs beta decay.  The results place $68\%$ confidence level upper limit on the cross section ranging from 0.1 barns to 0.005 barns(Eq.\ref{crosssection1}), both of which are on the order of strong interaction cross sections.\\ 
\begin{figure}[ht!]
\begin{center}
\includegraphics[scale=.42]{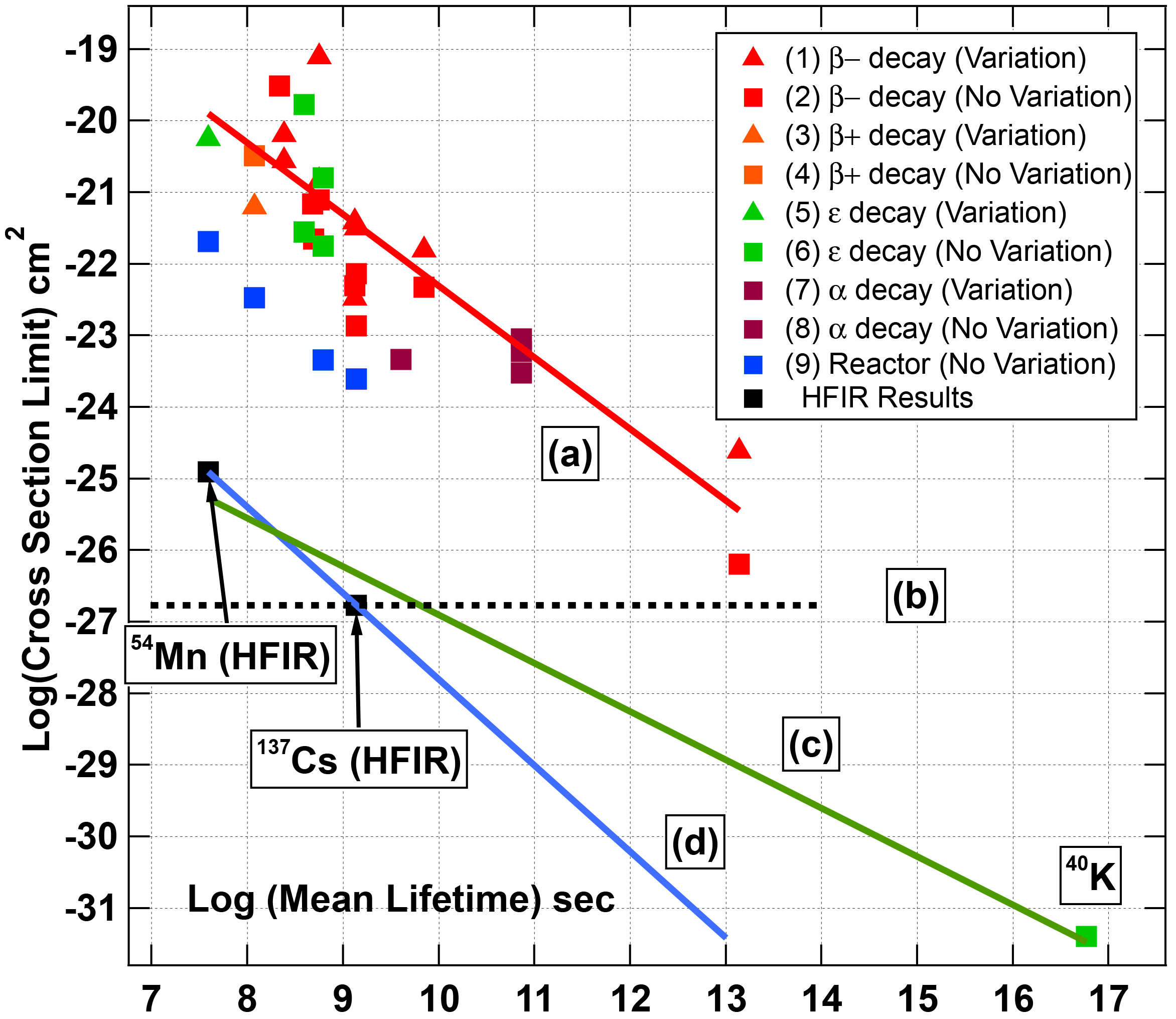} 
\end{center}
\caption{The logarithm cross section sensitivity of those experiments measuring, time-dependent variations, no evidence of variation, and these HFIR results, all as a function of the logarithm mean nuclear decay lifetime. The data are based on Table~\ref{limit_yes}, and \ref{limit_no}. The data includes (1) $\beta^{-}$ decay with variation results\protect\cite{Falkenberg200241,Veprev201226,ALBURGER1986168,Jenkins201281,parkhomov2005bursts,Baurov2007,Parkhomov2011,Sturrock2012755,Sturrock20168}, (2) $\beta^{-}$ decay with no effect results\protect\cite{Bruhn200228,Semkow2009415,Kossert201433,Kossert201518,Schrader20101583,Bellotti2013116,Siegert19981397}, (3) $\beta^{+}$ decay with variation results\protect\cite{OKeefe2013}, (4) $\beta^{+}$ decay with no effect results\protect\cite{Norman2009135}, (5) Electron Capture decay with variation results\protect\cite{Jenkins2009407}, (6)  Electron Capture decay with no effect results\protect\cite{Bellotti2013116, Schrader20101583, Siegert19981397}, (7) $\alpha$ decay with with variation results\cite{Jenkins200942}, (8)$\alpha$ decay with no effect results\protect\cite{Siegert19981397,Semkow2009415,Cooper2009267}, and (9) reactor antinuetrino as a test source with no effect  results\protect\cite{deMeijer2011320}. (10) The HFIR $^{54}$Mn and $^{137}$Cs  results. See the text for explanation of the curves (a), (b), (c), and (d).}
\label{Crosssection_Lifetime1}
\end{figure}
\indent
Figure~\ref{Crosssection_Lifetime1} compares the data available in the literature as displayed in Table \ref{limit_yes} and \ref{limit_no}, and this experiments final results in Table \ref{Final_Upperlimit_M} and \ref{Final_Upperlimit}. Each experiment is displayed by the logarithm of its cross-section or sensitivity found using Eq. \ref{crosssection1}, as a function of the logarithm of the mean lifetime (seconds). All experiments are included regardless of reporting observed variation or null observations. The reported decay modes include (1) the negative $\beta$-decay with time-dependent variation results (Red Triangles)\cite{Falkenberg200241,Veprev201226,ALBURGER1986168,Jenkins201281,parkhomov2005bursts,Baurov2007,Parkhomov2011,Sturrock2012755,Sturrock20168}, (2)the negative $\beta$-decay with "Null" variation results (Red Squares)\cite{Bruhn200228,Semkow2009415,Kossert201433,Kossert201518,Schrader20101583,Bellotti2013116,Siegert19981397} (3) the positive $\beta$-decay with time-dependent variation results (Orange Triangles)\cite{OKeefe2013} (4)the positive $\beta$-decay with "Null" variation results (Orange Squares)\cite{Norman2009135} (5) the electron capture $\beta$-decay with time-dependent variation results (Green Triangles)\cite{Jenkins2009407} (6) the electron capture $\beta$-decay with "Null" variation results (Green Squares)\cite{Bellotti2013116, Schrader20101583, Siegert19981397} (7) the $\alpha$-decay with time-dependent variation results (Purple Triangles)\cite{Jenkins200942} (8) the $\alpha$-decay with "Null" variation results (Purple Squares)\cite{Siegert19981397,Semkow2009415,Cooper2009267}, are by exposing the solar neutrino. (9) The reactor antineutrino "Null" variation results \cite{deMeijer2011320}. (10) The HFIR $^{54}$Mn experiment result. \\
\indent
In this search the meaning of an excluded region is not well defined.  However, it is expected that the cross section sensitivities for neutrino, and antineutrino interactions at a fixed measuring sensitivity should follow the curve,
\begin{equation}
\sigma=A \, \tau ^{P},
\end{equation}
where $A$, and $P$ are the fitting coefficients. As a  comparison a fit is made to those experiments reporting decay rate parameter variations and displayed in Figure~\ref{Crosssection_Lifetime1} as curve (a) and  in Table \ref{Cross_Fit}. Another approach is to assume that the cross section is the fundamental.  In this case Curve (b) compares this experiments $^{137}$Cs limit, its most sensitive cross section limit, to all experiments.  This experiment is more sensitive than all previous experiments reporting positive decay rate parameter variations, and thus it is in disagreement with all positive result experiments on this basis by a factor of 10$^4$. Curve (c) is a fit of this experiments two results and a null solar neutrino based experiment using $^{40}$K\cite{Bellotti2013116}. The connection between these two independent experiments results maps out an exclusion zone in the temporal cross-section space excluding decay rate parameter variations, again at a level $10^{4}$ times more sensitive than any previously reported positive result. Curve (d) displays the temporal cross exclusion zone if extrapolated using only this experiments two results, and again is in disagreement with all positive result experiments.\\
\begin{table}[t!]
\begin{center}   
%\setstretch{1.25}
\caption{Coefficients from four type of curve fitting with previous experiments.}
\begin{tabular}[b]{ccc}
\hline \hline
      Curve & $A$ & $P$    \\ 
        \hline
       (a) Observed variation & -12.31 & -1 \\
       (b) Cross section only comparison & -24.87& 0 \\
       (c) Exclusion Region & -19.54& -0.71\\
       (d) This experiment temporal exclusion & -17.28 & -1 \\
        \hline\hline        
        \end{tabular}       
\end{center}
\label{Cross_Fit}
\end{table}
\indent
The properties of these curves as displayed in Figure \ref{Crosssection_Lifetime1} and in Table \ref{Cross_Fit}, make a convincing case that those measurements reporting decay rate parameter variations are not consistent with the source of the variations being caused by neutrino or antineutrino interactions. 

\section{Acknowledgments}

The authors thank the HIFR staff for their operational support at Oak Ridge National Laboratory, necessary to accomplish the experiment.  Special thanks is given to John Herczeg, Nuclear Science Division of the Department of Energy for his support, encouragement, and interest in seeing this experiment to completion. This work was funded under a grant from the US DOE Office of Nuclear Energy, Contract No. DE-DT-000-4091.

\bibliography{PRC_Paper.bib}

\end{document}